\documentclass[proof]{WileyASNA-v1}

\articletype{ORIGINAL ARTICLE}%

\received{xx xxxx xx}
\revised{xx xxxx xx}
\accepted{xx xxxx xx}

\begin{document}

\title{Orbital parameters of the bright binary $\alpha$ Draconis based on amateur spectra from the STAROS database}

% List of institutions
\address[1]{Single Tracking Astronomical Repository for Open Spectroscopy (STAROS), \orgaddress{06600 Antibes,  \country{France}}}
\address[2]{Société Astronomique de France, \orgaddress{3 rue Beethoven, 75016 Paris,  \country{France}}}
\address[3]{Southern Spectroscopic Project Observatory Team (2SPOT), \orgaddress{45 Chemin du Lac, 38690 Châbons,  \country{France}}}
\address[4]{Université Bretagne Sud, \orgaddress{11 Rue André Lwoff, 56000 Vannes, \country{France}}}
\address[5]{Three Hills Observatory, \orgaddress{The Birches CA71JF, \country{UK}}}

\author[1,2]{G. Bertrand}
\author[1]{C. Buil}
\author[1,2,4]{M. Le Lain}
\author[1]{V. Desnoux}
\author[1,3]{O. Garde}
\author[1]{E. Bertrand}
\author[1]{A. Blais}
\author[1]{JJ. Broussat}
\author[1]{E. Bryssinck}
\author[1]{L. Dalbin}
% \author[1]{R. Diz}
\author[1]{X. Dupont}
\author[1]{A. Garrigós}
% \author[1]{M. Larsson}
\author[1]{M. Di Lazzaro}
\author[1,5]{R. Leadbeater}
\author[1]{V. Lecocq}
\author[1]{J. Lecomte}
\author[1]{A. Leduc}
\author[1]{P. Louis}
\author[1]{A. Maetz}
\author[1]{L. Ribé de Pont}
\author[1]{A. Stiewing}
\author[1]{S. de Visscher}
\author[1]{F. Weil}

\authormark{G. Bertrand \textsc{et al}}

\corres{\email{team@staros-projects.org}}

\abstract{We present the results of a STAROS monitoring campaign on the bright eclipsing binary star $\alpha$ Draconis. Over 200 high-resolution spectra were obtained by an international group of amateur astronomers. 
We were able to calculate an homogeneously covered radial velocity curve and redetermine the orbital elements of the $\alpha$ Dra system. From the data set, we estimated the quality of our observations and accuracy of our radial velocity measurements. Finally, we have also implemented a spectral disentangling method to search for the signature of a companion star and used atmosphere
models to constrain the atmospheric parameters of the system.}


\maketitle

\section{Introduction}\label{sec1}

$\alpha$ Draconis (Thuban, HD123299) is a well-studied spectroscopic binary of spectral type A0III with an orbital period of 51.41891 ± 0.00011 days \citep{Pavlovski2022}. The binary system was initially considered as an SB1 \textit{(Single-lined Spectroscopic Binary)} as only the spectrum of the primary star
was detected in the early observations. The first determination of the 
orbital parameters of $\alpha$ Dra was carried out by \citet{harper1907}, then refined with 
more recent studies~: \citet{kallinger2004}, \citet{bischoff2017}. The star was also first resolved with the Navy Precision Optical Interferometer (NPOI) by \citet{Hutter_2016}.

In 2019, \citeauthor{Bedding2019}, discovered by analyzing light curves from the TESS satellite
that $\alpha$ Dra is also an eclipsing binary, one of the brightest currently known
($V=3.68$). In 2022, \citeauthor{Pavlovski2022}, published a first analysis of the binary as an SB2 system (\textit{Double-lined Spectroscopic Binary}). They used the high-resolution spectrograph ($R=85000$) HERMES, installed on the Mercator telescope (1.2 m) in La Palma and data from the TESS satellite. \citet{Hey2022} presented an analysis of the eclipsing spectroscopic binary system. They obtained the parameters of this system by simultaneously adjusting the TESS light curve and the radial velocities (RV) acquired using the SONG spectrograph.

In this paper, we present the result of a new observation campaign of $\alpha$ Dra. It was built as a coordinated observation exclusively among amateur astronomers equipped with high resolution spectrographs. The observations were centralized, merged and analyzed using a new tool called STAROS ("Single Tracking Astronomical Repository for Open Spectroscopy"). The main goal of this work was  to evaluate the quality of amateur data in the context of observations usually performed by professional astronomers. We also wanted to demonstrate that the involvement of a large number of observers, unprecedented in the history of spectroscopy, can lead to very convincing analyses and results. All the data from this observation campaign are publicly available on the STAROS platform.\footnote{\href{https://alphadra.staros-projects.org}{https://alphadra.staros-projects.org}} 

\section{Spectroscopic Observations and data quality}\label{sec2}

\subsection{Observations}

A total of 25 individual observers from 9 countries took part in the campaign. 227 spectra were obtained, representing a total exposure time of 255 hours. The campaign ran from March 2023 to August 2023, which spans 3.5 orbital periods of the $\alpha$ Dra system.

The list of observers, together with the brief description of the instruments used is presented in Table~\ref{table:obstab}. Next to each name, a brief description of the instrument used is provided. The spectrographs used are mainly the Lhires III (produced by the Shelyak company) model \footnote{\href{https://www.shelyak.com/produit/lhires-iii/?lang=en}{https://www.shelyak.com/produit/lhires-iii/?lang=en}} and the Star'Ex model \footnote{\href{http://www.astrosurf.com/solex/sol-ex-presentation-en.html}{http://www.astrosurf.com/solex/sol-ex-presentation-en.html}}, a versatile spectrograph, self-built by the observers using 3D printing, giving a resolution range of between 1000 and 40000, depending on the configuration.

Telescope diameters range from 60 mm to 400 mm, with a variety of optical formulas (refractor, Newton, Schmidt-Cassegrain, Maksutov, etc.). The association of these instruments with the spectrographs provide spectral resolution R between 10000 and 25000. The spectra obtained exhibit a signal-to-noise ratio ($S/N$) of between 100 and 480 (measured between wavelengths of 6630 and 6650Å). The mean standard deviation of the calibration on all the reduced data is 0.014Å (range between 0.006Å and  0.028Å). 

%%%%%%%%%%%%%%%%%%%%%%%%%%%%%  TABLEAU 1 %%%%%%%%%%%%%%%%%%%%

\begin{center}
\begin{table*}[t]%
\scriptsize
  \caption{List of observers}
\tabcolsep=0pt%
\begin{tabular*}{500pt}{@{\extracolsep\fill}lllllcc}
\toprule
  \textbf{Name} & \textbf{Country} & \textbf{Telescope} & \textbf{Spectrograph} & \textbf{Camera} & \textbf{Resolution} & \textbf{Diameter} (mm)\\
\midrule
Alain Maetz & France & Mewlon 210  & LHIRES III &  Atik 460Ex & 16500 & 210 \\ 
  Albert Stiewing & USA & Celestron C14 &  LHIRES III& Atik 460Ex & 16200 & 355 \\
Antoine Blais &  France &  SkyWatcher 80ED & Star'Ex&Zwo ASI 290MM & 14000 & 80\\
Antonio Garrigós & Spain& Celestron C11&LHIRES III& Atik 460Ex & 14500 & 279 \\ 
Arthur Leduc& France& Newton 130F8 & Star'Ex& Zwo ASI 183MM PRO& 15000 & 130\\
Arthur Leduc& France& Newton 200F5&Star'Ex&  Zwo ASI 183MM PRO& 13000 & 200\\
Arthur Leduc& France& MTO 100F7.5&Star'Ex&  Zwo ASI 183MM PRO & 15000 &100\\
Christian Buil& France& Askar 107 PHQ& Star'Ex&  Zwo ASI 533MM PRO & 19000 &107\\
Christian Buil& France& Askar 80 PHQ& Star'Ex&  Zwo ASI 533MM PRO& 19500 & 80\\
Christian Buil& France& 100 ED&Star'Ex& Zwo ASI 533MM PRO& 18000 & 100\\
Christian Buil& France& FSQ106ED&Star'Ex&  Zwo ASI 533MM PRO & 18000 &106\\
Christian Buil& France& ASKAR 300&Star'Ex&  Zwo ASI 533MM PRO & 21000 &60\\
Christian Buil& France& Newton 200&Star'Ex& Zwo ASI 533MM PRO& 19000 & 200\\
Erik Bryssinck& Belgium& Celestron C11&LHIRES III&  Starlight Xpress Trius-SX814   & 17000 &279\\
Etienne Bertrand& France& Celestron C11&LHIRES III& Atik 314L+& 17000 & 279\\
Franck Weil& France& Maksutov 127&Star'Ex&Zwo ASI 183MM PRO& 12000  &127\\
Guillaume Bertrand& France& SkyWatcher 72ED&Star'Ex&Zwo ASI 183MM PRO& 21000  &72\\
Jean-Jacques Broussat& France& Celestron C9& LHIRES III& Atik 460& 17000  &279\\
Julien Lecomte& USA& AT130EDT& Star'Ex&Zwo ASI 533MM PRO& 17000  &130\\
Laurent Dalbin& France& Refractor 76/432& Star'Ex&  Zwo ASI 178MM& 12000 & 76\\
Luis Ribé de Pont& Spain& Celectron C11& LHIRES III& Atik 460 EX& 21600 & 279\\
Magnus Larsson& Sweden& Celestron C8& Star'Ex& QHY183& 13000 & 203\\
Massimo Di Lazzaro& Italy& Celestron C8HD&Star'Ex& Starlight Xpress Trius-SX825& 10000 & 203\\
Matthieu Le Lain& France& Mak127&Star'Ex& Zwo ASI 183MM PRO& 13300 & 127\\
Olivier Garde& France&RC400 Astrosib& Whoppshel (ESP)&Atik 460EX& 25000 & 400\\
Pascal Louis& France& SkyWatcher 80ED& Star'Ex& Zwo ASI 183MM PRO& 13000 & 80\\
Rick Diz& USA& Celestron C8&Star'Ex& Zwo ASI 183MM PRO& 14000 & 203\\
Robin Leadbeater& UK&Celestron C11& LHIRES III& Atik 460& 17000 & 279\\
Simon de Visscher&  Switzerland&AP130EDT& LHIRES III& Zwo ASI 533MM PRO& 15000 & 130\\
Valerie Desnoux& France&ASKAR107&Star'Ex& Zwo ASI 183MM PRO& 20500 & 107\\
Vincent Lecocq& France& FSQ106@F8&Star'Ex& Zwo ASI 183MM PRO& 22000 & 106\\
Vincent Lecocq& France&  M703@F10 &Star'Ex&Zwo ASI 183MM PRO& 14000 & 180\\
Xavier Dupont& France& Newton 115x900&Star'Ex&Zwo ASI 183MM PRO& 19300 & 115\\
Xavier Dupont& France& Dall-Kirkham 350@F5.5&Star'Ex&Zwo ASI 183MM PRO& 21000 & 350\\
\bottomrule
 \end{tabular*}
    \label{table:obstab}
\end{table*}
\end{center}

%%%%%%%%%%%%%%%%%%%%%%%%%% END TABLEAU 1 %%%%%%%%%%%%%%%%%%%

\subsection{Quality of spectra, measurement precision}

In the context of the $\alpha$ Dra campaign, the most important quality and performance criterion is the accuracy of radial velocity measurement. The aim is to maximize this accuracy and therefore minimize measurement error. 

The quality of the radial velocity measurement from a high resolution spectrum depends on several factors. The dominant ones are the noise present in the raw data, the width and contrast of the spectral lines of interest, and the number of lines which are actually used. These parameters alone enable us to obtain, by calculation, a rough estimate of the expected measurement accuracy (in a so-called "photon noise" regime, the only one considered in this study). The procedure for achieving this result is described below.

However, it's essential to realize that measurement error, which can be subdivided into noise on the one hand and bias on the other (representing random and systematic error respectively), is linked to many other factors, e.g. mechanical deformations of the spectrograph during observation, its thermo-elastic variations, the way the entrance slit is illuminated, the technique chosen to calibrate the spectrum in terms of wavelength. All these elements have an impact on the accuracy of radial velocity measurement. We refer to these elements as "pseudo-noise" to distinguish them from detection noise. While the latter can be estimated fairly easily using a statistical approach, the operation becomes much more complex when it comes to pseudo-noise (for example, some of these noises are additive, others compensate for each other.). The most reasonable method of judging their effect is to compare observations with theoretical results, and/or to compare results obtained by many other observers. The procedure we adopted is described below.

The measurement error in characteristic radial velocity when a single line in the spectrum is exploited and in the photon noise regime (detector noise is considered negligible, as are other noise sources) is given by a classic formula from \citet{murphy2007}
\begin{equation}
{\delta\nu = A \cdot \frac{\mathrm{FWHM}}{C\ \cdot S/N\ \cdot \sqrt{n}}}
\end{equation}

with, $A$, a coefficient that depends on the shape of the line, $\mathrm{FWHM}$, the full width at half maximum of the line, $C$, the contrast of the line (the ratio between its depth and the continuum level), the $S/N$ ratio in the continuum and $n$, the number of sample elements (pixels) in the $\mathrm{FWHM}$ width (to be measured in a raw spectral profile).

\begin{figure}    
     \includegraphics[width=\linewidth]{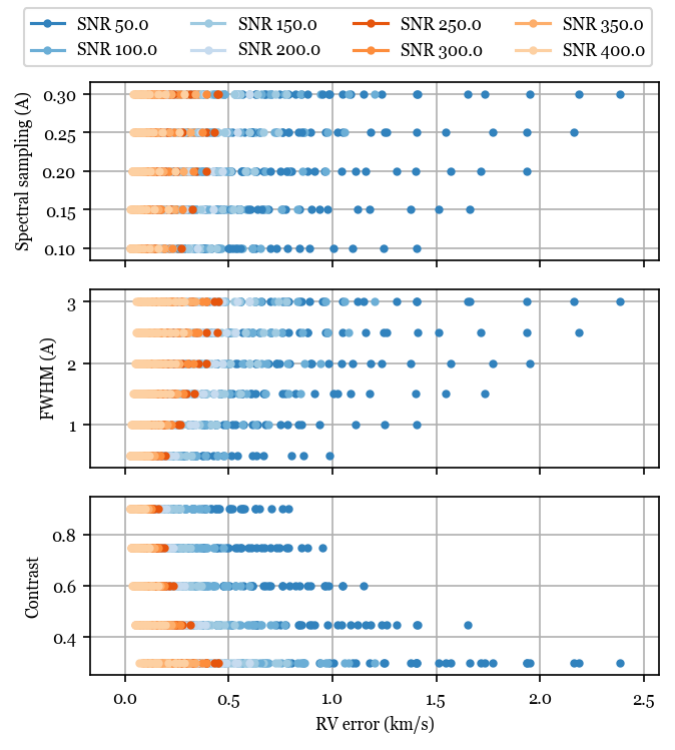}
       \caption{$\delta\nu$ error in km s$^{-1}$ from Monte Carlo simulation as a function of spectral sampling ($n$), FWHM and the contrast of the line($C$). Each point represents the standard deviation of the Gaussian fit over 1000 synthetic spectra.\label{fig:montecarlo}}
\end{figure}

The values of $A$ coefficient vary in the published literature. We conducted our own analysis and carried out multi-parameter Monte Carlo simulations (See Figure \ref{fig:montecarlo}). Here are the values obtained for three different fitting models~:\\
 
\noindent Gaussian profile~: $A = 0.815$ ± $0.018$\\
Voigt profile~: $A = 0.862$ ± $0.038 $\\
Lorentz profile~: $A = 0.939$ ± $0.043 $\\ 

The $\mathrm{FWHM}$ must be expressed in km s$^{-1}$. From a measurement in Angstroms, we obtain the value in km s$^{-1}$ by doing the following~:
\begin{equation} 
\mathrm{FWHM}_{km s^{-1}} = c\ \cdot \frac{\mathrm{FWHM}_{A}}{\lambda}
\end{equation}
\noindent with $c$, the speed of light in vacuum and $\lambda$, the wavelength in Angstroms.

If $p$ is the spectral sampling value in km s$^{-1}$, and noting that $n = \mathrm{FWHM} / p$, we find another form of error calculation, sometimes encountered in the literature~:
\begin{equation}
{\delta\nu = A \cdot \frac{\sqrt{\mathrm{FWHM}} \, \sqrt{p}}{C\ \cdot S/N}}
\end{equation}

Unsurprisingly, line position measurement is most accurate when the line is narrow (corresponding to high spectral resolution and/or a star with naturally fine lines). It's also crucial that the line has a high contrast. Moreover, accuracy increases in proportion to the $S/N$ ratio measured in the continuum, as discussed below. It should also be noted that the spectrum must be correctly sampled to cover the interval equivalent to the width of a line with a significant number of detector pixels during acquisition (At least two for respect Nyquist critera).

\begin{figure}    
     \includegraphics[width=\linewidth]{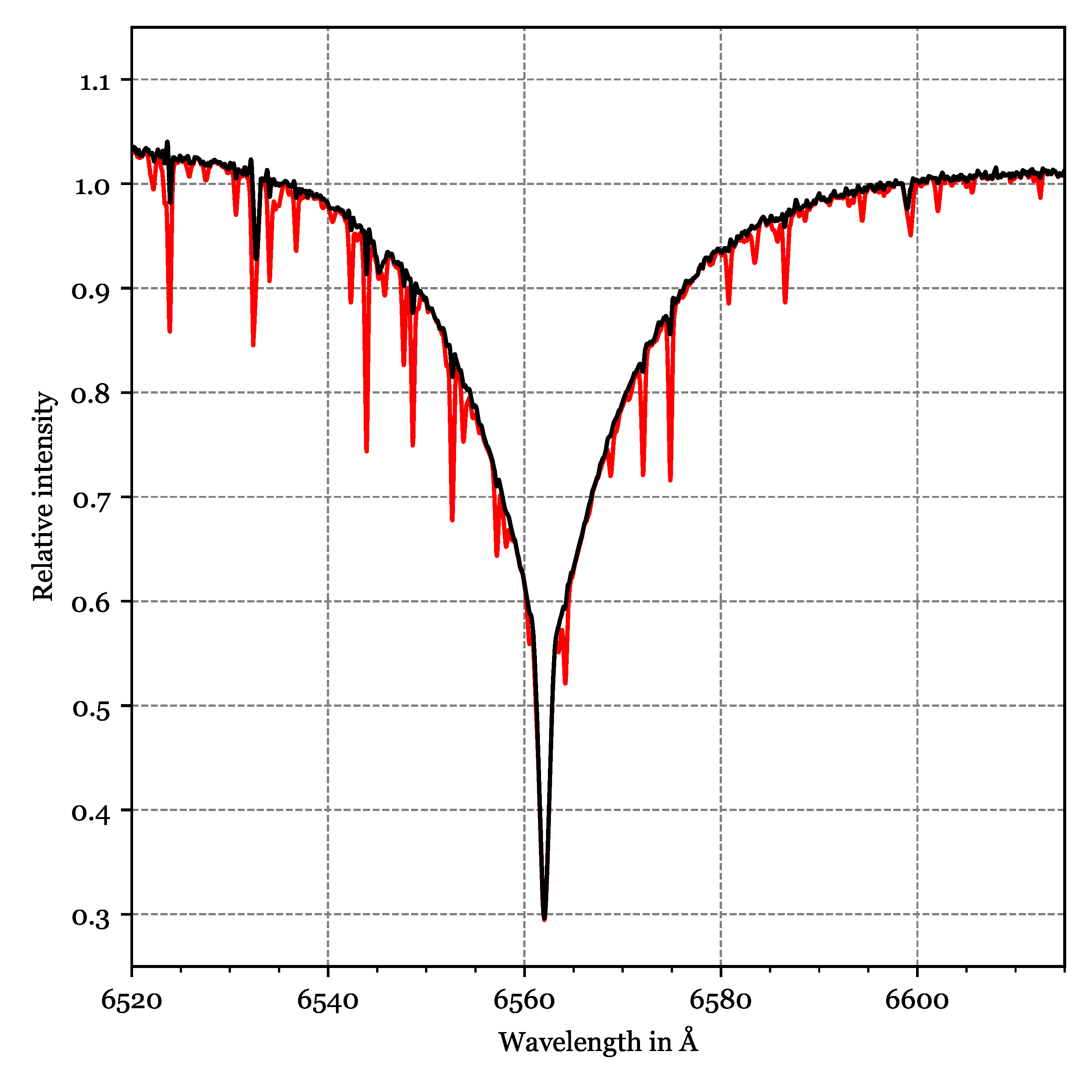}
       \caption{Result of the automatic removal algorithm
    of telluric lines. In red, the original spectrum; in black, the spectrum after
    telluric line removal.    \label{fig:cbuiltel}}
\end{figure}

The spectral range chosen for this campaign is that which encompasses the H$\alpha$ line of hydrogen, located at a wavelength of 6562.82 Å. In the spectrum of the star $\alpha$ Dra, this line is very broad and displays a characteristic Voigt profile. Preliminary work on our part has shown that the accuracy of radial velocity measurements reaches its maximum when focusing on the center of this profile, where the line is narrowest and takes on the appearance of a Gaussian function. This decision makes it possible to bypass the fluctuations induced by telluric lines (H$_2$O), even if they are previously eliminated (See Figure \ref{fig:cbuiltel}) thanks to a molecular atmospheric transmission model systematically applied (there are always residues left by the shrinkage process). In addition, the spectral signature of the system's second companion is also minimized. While it be true, in this case, that this narrow region of the line profile is free of water lines, we point out that this may not be true in general, depending on the  intrinsic RV of the target, range of orbital velocities and the heliocentric correction. 

\begin{figure}
     \includegraphics[width=\linewidth]{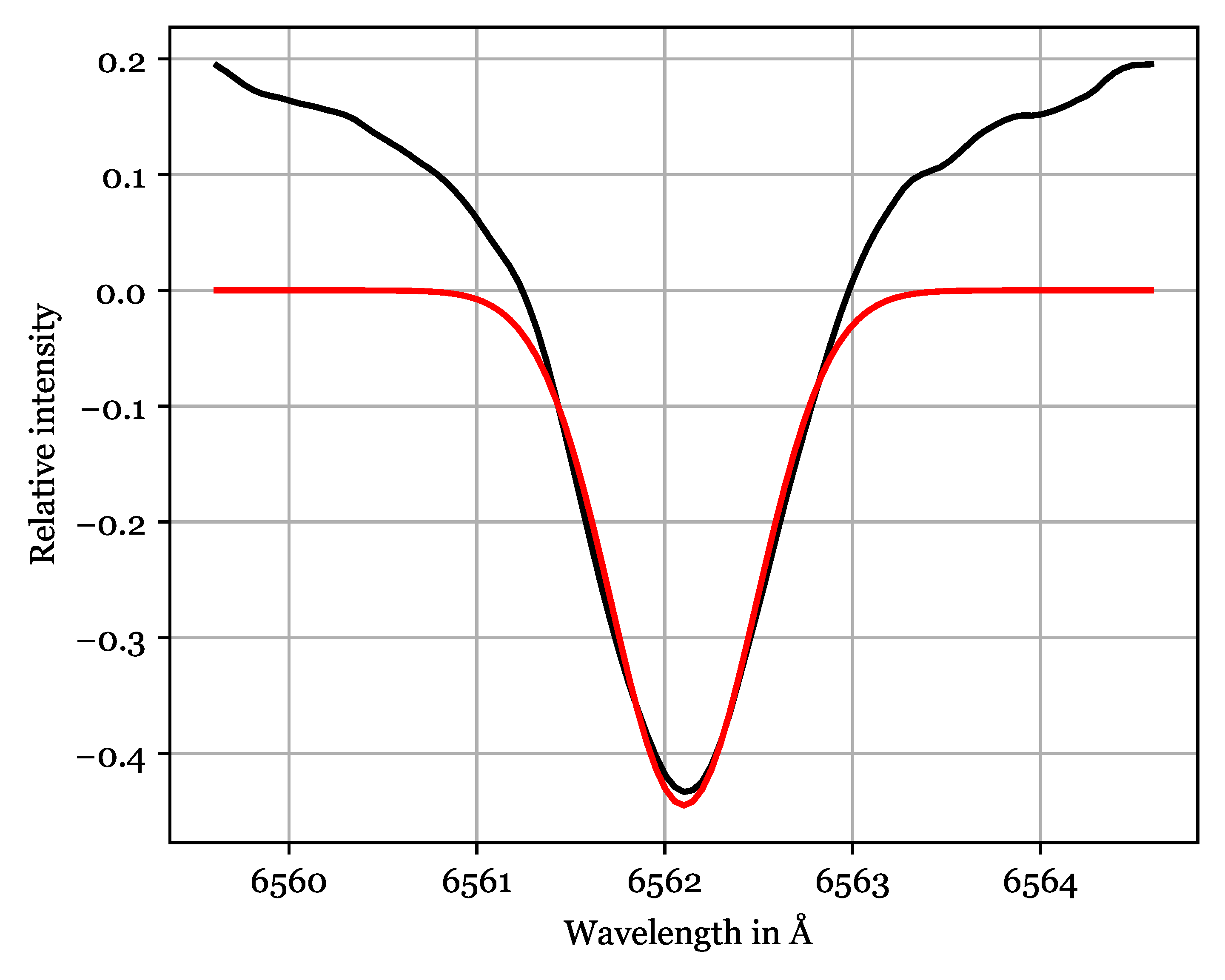}
    \caption{Area of least-squares adjustment of the line position, with the calculated profile in red and the observed profile in black. Portion of spectrum from 2023-04-24, C. Buil}
    \label{fig:cbuilfit}
\end{figure}

For illustration, we present here the analysis of the value of $\delta_v$ under the typical conditions of one of the participants (C. Buil), with a typical $S/N$ ratio of 350 measured on the continuum of observation data. For our calculation, it is necessary to correct this value, as we are not measuring the line against the actual continuum, but against a pseudo-continuum of lower intensity, corresponding to the plateau drawn in red in Figure~\ref{fig:cbuilfit}. If the true continuum is normalized to unity, it appears that the level of our pseudo-continuum is 0.48. Consequently, we adjust the $S/N$ to $S/N = 0.48 \times 350 = 168$.

The FWHM of the line's core is about 1 Å for our star, which translates into a FWHM expressed in velocity of~: 
\begin{equation}
\mathrm{FWHM} = 299 792.458 \times 1.0 / 6563 \approx 46\ km s^{-1}
\end{equation}

Note that the line, including its Gaussian core, is well resolved by the spectrograph used, a Star'Ex, which works here at a resolving power of $R=19000$, i.e. a spectral sharpness at the level of the hydrogen red line of $0.34$Å. The $\mathrm{FWHM}$ is sampled by 7 pixels, hence $n$ = 7. Relative to the pseudo-continuum, the contrast of the line is measured at $C = 0.42$ (if $S1$ is the intensity taken at the center of the line and $S0$ is the continuum intensity, then $C$ = 1 - $S1$/$S0$). The theoretical RV error will be~:

%\begin{equation}
%\label{rv_err_eq}
%\begin{split}
%\delta\nu = A \times \frac{\mathrm{FWHM}}{C\ \times S/N \dot \sqrt{n}}
%\\ 
%\delta\nu =  0.815 \times \frac{46}{0.42 \times  168 \times \sqrt{7}} = %0.20 \ \mathrm{km\ s}^{-1}
%\end{split}
%\end{equation}

\begin{equation}
\label{rv_err_eq}
\begin{split}
\delta\nu &= A \times \frac{\mathrm{FWHM}}{C \times (S/N) \times \sqrt{n}} \\
          &= 0.815 \times \frac{46}{0.42 \times 168 \times \sqrt{7}} \\
          &\approx 0.20 \ \mathrm{km\ s}^{-1}
\end{split}
\end{equation}

The calculated error corresponds to the best possible result under the defined conditions. In practice, the actual measurement is always less accurate. For example, for the observer considered in the example, the O-C RMS statistical error measured from numerous observations is 0.31 km s$^{-1}$. The disparity between theory and reality can be explained by the presence of the many pseudo-noise sources mentioned above, which have not been taken into account in our simplified formulation.

In fact, one of the important results we've drawn from the $\alpha$ Dra STAROS campaign is precisely the impact of noise in the data (more specifically the $S/N$ ratio), which enters directly into the calculation of radial velocity accuracy (\ref{rv_err_eq}).

It is therefore ideal to get as close as possible to the photon noise regime (the ultimate quality). We have quantified this deviation for each observer who submitted a sufficiently significant number of spectra for this campaign.

We assume that the signal measured, denoted $S$, is proportional to the collecting surface of the telescope used and the exposure time accumulated to produce the final spectrum of our star. The collecting surface is itself proportional to the square of the telescope's diameter, resulting in a pseudo-estimation of the collected signal~:
\begin{equation}
{S' \approx D^{2} \, t}
\end{equation}

It is obvious that in the ideal situation of the photon noise regime during observation, the $S/N$ ratio is proportional to the square root of the signal, hence, for our simulation, the following expression for the pseudo $S/N$ ratio~:
\begin{equation}
{S/N' \propto \sqrt{S'} = D\,\sqrt{t}}
\end{equation}

On this basis, we establish a quality criterion, $Q$. For each observer (and therefore specific instrumentation) and each observation of the star whose spectrum is subjected to STAROS, we evaluate the ratio~:
\begin{equation}
{Q = \frac{S/N_{obs}}{S/N'}}
\end{equation}

The observed $S/N$ ratio, $S/N_{obs}$, is measured in the continuum, in a spectral line-free zone.

The units are arbitrary, since the important thing is to establish the relative value of $Q$ between observers. For example, if we use a telescope 107 mm in diameter and the exposure time is 30 minutes, then,
\begin{equation}
{S/N' = D\sqrt{t} = 10.7 \times \sqrt{0.5} \approx 7.57}
\end{equation}

The higher the value of $Q$, the better the performance of the instrument. In theory, the dispersion of radial velocity measurements should also improve (excluding calibration bias). On the other hand, a low value for $Q$ indicates a potential problem, very often linked to an inappropriate slit width for the telescope used, a failure to focus the star on the spectrograph's entrance slit, or a poor centering of the star's image on the same slit - these are the most commonly encountered problems. Table~\ref{table:precision_vr} summarizes the situation for some of the campaign's participants, whose number of observations satisfies the requirements of a statistical analysis (at least 8 spectra).

\begin{center}
\begin{table}[t]%
\centering
   \caption{Estimation of the quality of spectra.\label{table:precision_vr}}%
\tabcolsep=0pt%
\begin{tabular*}{20pc}{@{\extracolsep\fill}lcccc@{\extracolsep\fill}}
\toprule
    \textbf{Observer} &\textbf{Obs. Count} & \textbf{Q} & \textbf{Std O-C} &  \textbf{Std Calib} \\
\midrule
    C. Buil & 39 & 37.4 & 0.31 & 0.44 \\
    V. Lecocq & 16 & 31.2 & 0.78 & 0.61 \\
    A. Leduc & 19 & 22.0 & 0.71 & 0.99 \\
    S. de Visscher& 8& 19.7& 0.95& 0.46 \\
    E. Bryssinck& 16& 19.7& 0.64& 0.30 \\
    G. Bertrand& 20& 18.2& 1.03& 0.68 \\
    A. Stiewing& 9& 18.2& 1.06& 0.51 \\
    E. Bertrand& 10& 17.2& 0.58& 0.57 \\
    X. Dupont& 35& 12.6& 1.07& 0.59 \\
    A. Garrigos& 8& 8.1& 0.70& 0.61 \\
    R. Diz& 8& 6.8& 1.53& 1,30 \\
\bottomrule
\end{tabular*}
\begin{tablenotes}
\item Std O-C and Std Calib are expressed in km s$^{-1}$
\end{tablenotes}
\end{table}
\end{center}

For every observer, we list the number of observations made, the value of the parameter $Q$, the dispersion of measurements (standard deviation between the observed radial velocities and the radial velocities calculated from the best orbit found for the binary), and finally the mean deviation between the observed velocities and the velocities calculated from our orbit model.

 Our idea is to compare the value of the parameter $Q$ and the dispersion of each observer's measurements ("Std O-C" column). Although the correlation is not totally obvious, a trend emerges~: the higher the value of $Q$, the more accurate the radial velocity measurement, which is in line with expectations.

It is important to note that this exercise is not straightforward and remains partially incomplete at this stage. The $S/N$ ratio in the continuum is sometimes difficult to measure, and we have not taken into account variations in spectral resolution between observers (the median value measured on telluric lines is $R=17500$ for this campaign, with a fairly small dispersion around this value). This preliminary work should be completed in future campaigns.

The "Std Calib" column provides an estimate of the spectral calibration error. This is the mean standard deviation calculated by systematically measuring the position of several telluric lines. 

Based solely on the hydrogen line profile of our target star, it is clear that we are able to obtain radial velocity measurements with an accuracy of around 1 km s$^{-1}$, and even slightly better.

\section{RADIAL VELOCITY MEASUREMENTS}

For RV measurements of $\alpha$ Dra, we determined the central wavelengths $\lambda$ of the Balmer lines H$\alpha$ ($\lambda_0$ = 6562.82 Å). The determination of line centers was performed with a user-friendly STAROS Python script\footnote{\textbf{SpectroBinaryStarSystem}~: Python 3 library for automatic measurement of radial velocity measurement on a series of spectra and format the result
(radial velocity curve, 2D spectrum etc.)
\href{https://github.com/guillbertrand/spectrobinarystarsystem}{https://github.com/guillbertrand/spectrobinarystarsystem}}, which includes the following steps : 

- Wavelenght calibration check by measuring tellucic lines.

- Telluric lines removal (See Figure \ref{fig:cbuiltel}).

- Barycentric correction.

- H$\alpha$ line centroid detection.

- Extraction of the spectral region of interest, 2.5 Å wide on either side of the centroid found.

- Intensity normalization of the extracted spectral region.

- Gaussian fitting ($\mathrm{FWHM}$ fixed at 1Å) on the core of the $H\alpha$ line (See Figure \ref{fig:cbuilfit}). The fitting method uses the Levenberg-Marquardt algorithm via the Python library Astropy~\citep{astropy2013}. This is the same method used above for the Monte Carlo simulation of the error $\delta\nu$.

- Calculation of the RV with the relation \ref{doppler} where c is the speed of light and BC the barycentric correction. 

Dynamical spectra of the H$\alpha$ line was computed (See Figure~\ref{fig:flux2d}).
The measured RVs for all the observation periods are listed in appendix~\ref{table:measurments_vr}.

\begin{equation}
\label{doppler}
{RV = c \cdot \frac{\lambda - \lambda_0}{\lambda_0} + BC}
\end{equation}

\begin{figure}
   
     \includegraphics[width=\linewidth]{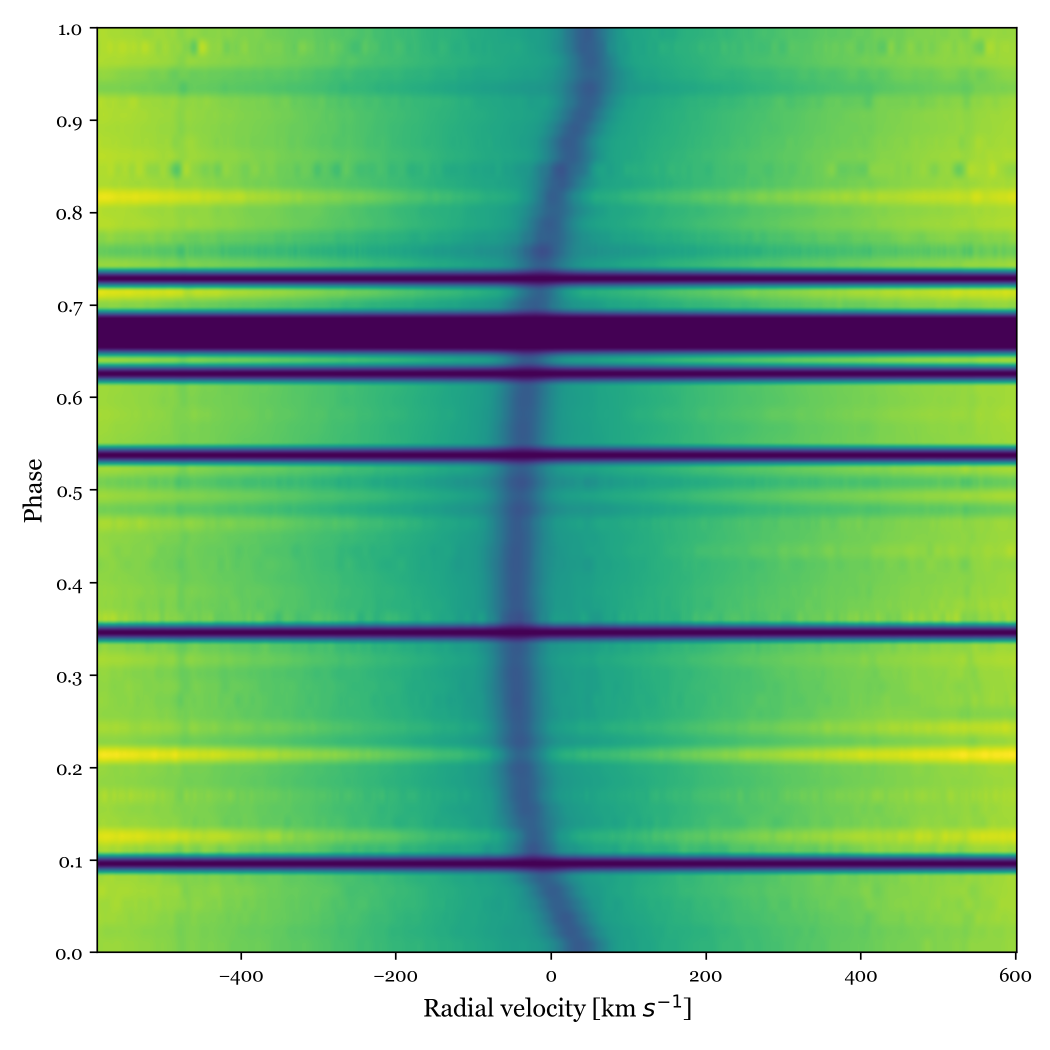}
      \caption{Dynamical spectra of the H$\alpha$ line in Doppler space.\label{fig:flux2d}}
\end{figure}

\section{ORBIT DETERMINATION}

The orbital elements, namely the true anomaly $\upsilon$, the periastron angle $\omega$, the semi-amplitude $K$, the eccentricity $e$, and the radial velocity of the center of mass $\gamma$ are related to the RV by the relation: 

\begin{equation}
\label{rv_equation}
{RV = \gamma + K \cdot [cos(\upsilon+ \omega) + e\ cos(\omega)]}
\end{equation}

The semi-amplitude $K$ depends on the minimum semi-major axis $asin(i)$, the orbital period $P$, and the orbital eccentricity $e$ of the component of the spectroscopic binary system.

\begin{equation}
\label{k_equation}
{K = \frac{2\pi \cdot a sin(i)}{P \cdot \sqrt{1-e^2}}}
\end{equation}

Orbital elements of $\alpha$ Dra were determined by fitting orbital solutions on our obtained RV data by using the BinaryStarSolver library~\footnote{
\textbf{BinaryStarSolver}~: Python 3 library for solving the orbital parameters of a binary system
SB1 or SB2 binary system from a radial velocity time series.
Barton, C. \& Milson, N. \href{https://arxiv.org/pdf/2011.13914.pdf}{https://arxiv.org/pdf/2011.13914.pdf}
\href{https://github.com/NickMilsonPhysics/BinaryStarSolver}{https://github.com/NickMilsonPhysics/BinaryStarSolver}} to find the period and 
orbital parameters. The radial velocity data are fitted to a curve using a nonlinear least-squares approach (Levenberg-Marquardt) to minimize residuals.

\begin{figure*}
\centering

     \includegraphics[width=1.0\textwidth]{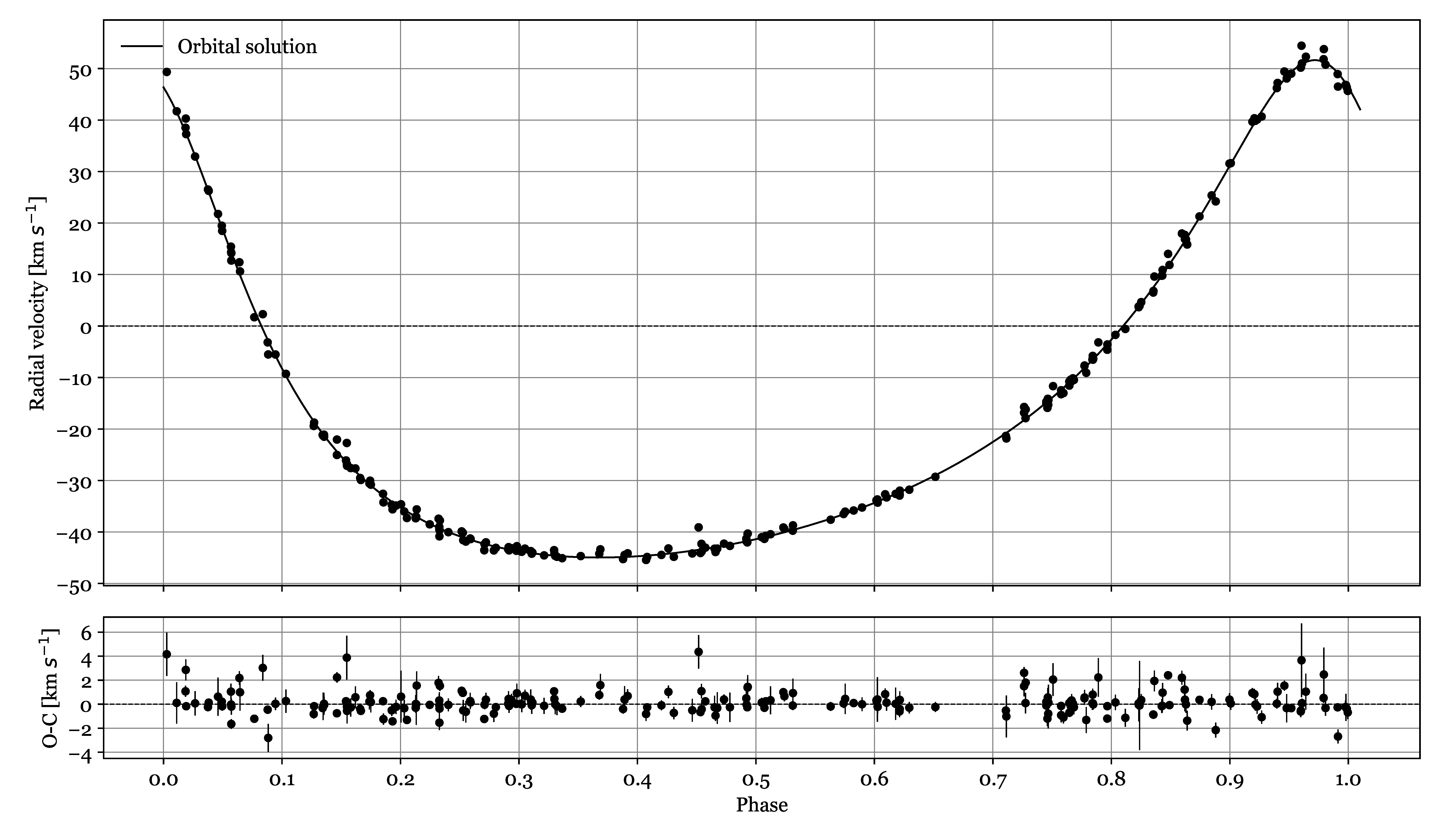}
\caption{Radial velocity curve calculated from 227 observations in the database
STAROS database. The uncertainties include the $\delta\nu$ error described in paragraph 2.2, and the calibration error calculated from carefully selected telluric lines around the H$\alpha$ line.\label{fig:rv}}

\end{figure*}

 The periastron epoch $T0$ was fixed at 2451441.804 \citep{Pavlovski2022}. The derived orbital solution is illustrated as black line in the phase-folded RV curve in Figure \ref{fig:rv}, and the orbital elements with their uncertainties are summarized in table \ref{table:orbital_parameters}.  Our solution is in agreement with previous works \cite{Hey2022, Pavlovski2022}.

\begin{table*} 
    \centering
   
        \caption{Orbital parameters calculated from 227 observations in the STAROS database.}
    \begin{tabular}{lcccc}
        \hline
        Parameter & Unit & R. Bischoff et al. 2017 & K. Pavlovski et al. 2022 & \textbf{STAROS 2023} \\
        \hline

Period $P$	 & [d]	 & 51.440 ± 0.024	 & 51.41891 (fix)	 & 51.4203 ± 0.0102 \\

Primary star's semi-amplitude $K$	 & [km s$^{-1}$]	 & 47.48 ± 0.21	 & 48.512 ± 0.054	 & 48.296 ± 0.075 \\
Systemic radial velocity $\gamma$	 & [km s$^{-1}$]	 & -13.50 ± 0.12	 & & 	-15.581 ± 0.043 \\
Periastron angle $\omega$  & [°]	 & 21.8 ± 0.6	 & 21.28 ± 0.13	 & 21.1235 ± 0.1845 \\
Eccentricity $e$		 &  & 0.426 ± 0.004	 & 0.4229 ± 0.0012 & 	0.41997 ± 0.00115 \\
Mass function $f(M)$	 & [M$\odot$]	 & 0.422 ± 0.008	 &  & 	0.44863 ± 0.00227 \\
Periastron epoch $T0$	 & [d]	 & 2457406.38 ± 0.09	 & 2451441.804 ±0.014	 & 2451441.804 (fix) \\
    \end{tabular}

 \label{table:orbital_parameters}
\end{table*}

\section{Secondary component extraction}

Spectral disentangling is an efficient method for extracting the spectra of individual components. To accomplish this, a time-series of high-resolution spectra is needed, preferably uniformly distributed orbital phase. \citet{Quintero2020} have carried out a thorough analysis of a spectral disentangling method based on a shift-and-add process in wavelength space. 

We have created a Python algorithm, largely inspired by the efficient and reliable disentangling method described by \citet{Quintero2020}. This method 
quickly reconstructs line profiles with correct fluxes and without artefact. To apply this, we need to determine the RV of the secondary star.

\citet{Pavlovski2022} claim $M_A$ = 3.186 M$\odot$ and i=85.4°, which we
will adopt. Using our mass function f(M) = 0.44863 and knowing that :

\begin{equation}
{f(M)=\frac{M_B^2 \cdot sin^3(i)}{(M_A +M_B)^2}}
\end{equation}

we find $M_B$ = 2.430M $\odot$ and a mass ratio q = $M_B$ / $M_A$ = 0.763.
This ratio allows us to determine the RV of the secondary component of the system using the formula below with $RV_A$ the measured radial velocity of the primary component.
\begin{equation}
RV_B = \gamma - \frac{(RV_A -  \gamma)}{q}
\end{equation}

\begin{figure*}

     \includegraphics[width=\linewidth]{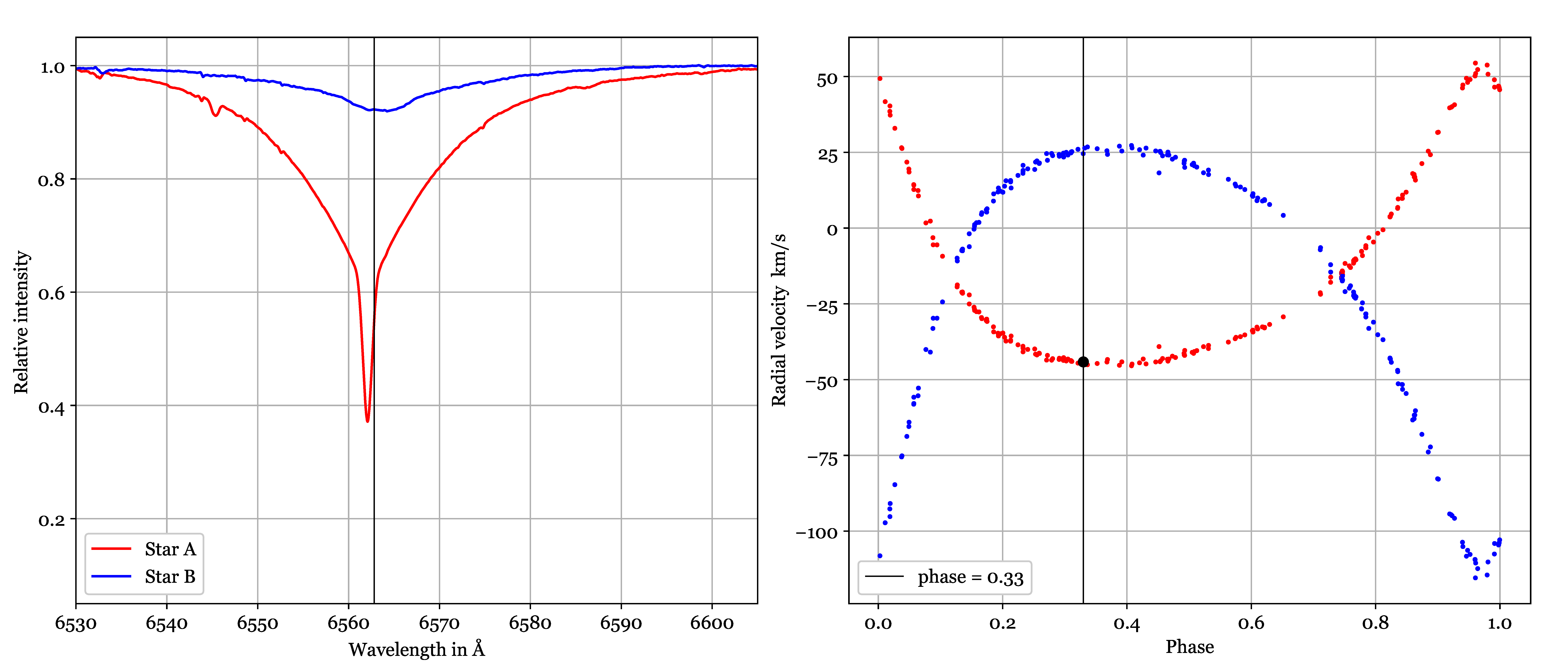}

\caption{Disentangled spectra of the primary (red line) and secondary (blue line) components of the $\alpha$ Dra system. Excerpt from \href{https://bit.ly/3Pw5fHL}{https://bit.ly/3Pw5fHL} animation.\label{fig:disentangling}}

\end{figure*}

We ran our script using 227 spectra from the STAROS database and the RVs of the two components. Our script converges after 1000 iterations and shows for the spectrum of the secondary component (See Figure \ref{fig:disentangling}) a weak absorption line significantly broadened by rotation, indicating that the secondary is rotating rapidly. We note that the H$\alpha$ line appears as a Voigt profile for the primary component and the presence (on the blue wing) of a weak absorption line (Mg II - 6545.97 Å).

\section{Atmospheric parameters}

\begin{figure}
     \includegraphics[width=\linewidth]{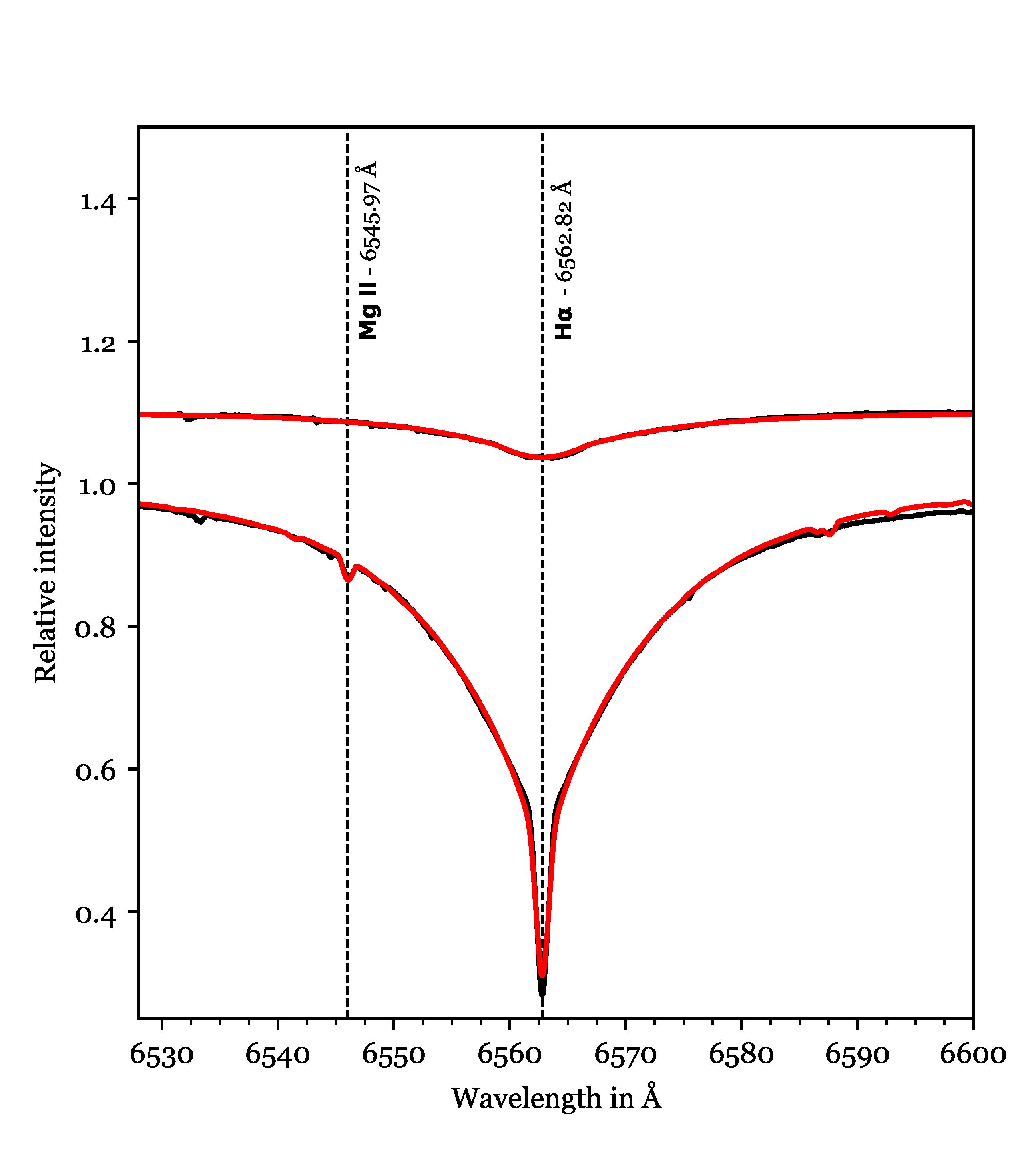}
     \caption{Quality of the fit (red) to the disentangled spectra (black) of the primary (bottom spectrum) and secondary (top spectrum) in H$\alpha$ region.\label{fig:atmofit}}
\end{figure}

We have adopted the plane-parallel, hydrostatic, line-blanketed local thermodynamic equilibrium (LTE) grids of ATLAS9 model atmospheres calculated by \citet{castelli2004new}. For the primary component we have computed a grid of spectra using the program SPECTRUM \citep{Gray1982} for the temperature range 8000 $\leq$ $T_{\mathrm{eff}}$ $\leq$ 13000 K, with step 250 K, surface gravity 2.5 $\leq$ log $g$ $\leq$ 5 dex, with step 0.25 dex and rotational velocity  15 $\leq$ $\nu$ sin \textit{i} $\leq$ 40 km s$^{-1}$, with step 1 km s$^{-1}$. We adopt the grid computed at turbulent velocity $\xi$ = 0 km s$^{-1}$ and metallicity [M/H]=0.0. All synthetic spectra are convolved with a Gaussian linespread function using the program SMOOTH2 \citep{Gray1982} to reduce the resolution of the synthetic spectra to the observed one. We used $\chi^2$ routine to compare the observed spectra with the theoretical one. The best fitted model spectra (See Figure \ref{fig:atmofit}) was achieved for the primary star with $T_{\mathrm{eff}}$ = 9250 ± 250 K, log $g$ = 3.5 ± 0.5 dex, $\nu$ sin \textit{i} = 27 ± 1 km s$^{-1}$. 

For the secondary component we have computed grid of spectra for the temperature range 8000 $\leq$ $T_{\mathrm{eff}}$ $\leq$ 13000 K, with step 250 K, surface gravity 2.5 $\leq$ log g $\leq$ with step 0.25 dex and rotational velocity  100 $\leq$ $\nu$ sin \textit{i} $\leq$ 200 with step 1 km s$^{-1}$. We adopt the grid computed at turbulent velocity $\xi$ = 0 km s$^{-1}$ and [M/H]=0.0. The best fitted model (See Figure \ref{fig:atmofit}) for the secondary star was achieved with $T_{\mathrm{eff}}$ = 11000 ± 250 K, log $g$ =4.0 ± 0.5 dex, $\nu$ sin \textit{i} = 170 ± 1 km s$^{-1}$.

As it is shown, there is a good agreement between the observed and model spectrum for Mg II (6545.97 Å) and H$\alpha$ (6562.82 Å) line. However, the reduced spectral range ($\approx$ 6510 to 6660 Å) limits the quality of the analysis.

\newpage
\section{Conclusions}

In this study, we demonstrate that it is possible to achieve a radial velocity measurement accuracy of less than 1 km s$^{-1}$ using spectral data acquired by amateur astronomers. The data are heterogeneous in terms of equipment, reduction process, $S/N$, exposure time and resolution, but all are very well calibrated ($\approx$ 0.014 Å). A rigorous analysis of the quality of the data and the accuracy of the line centre measurement was carried out before proceeding with the data analysis. 

We were able to accurately calculate the orbital parameters of the $\alpha$ Dra system using only the spectral data. The calculated parameters are in good agreement with recent studies (\citealt{Hey2022, Pavlovski2022}).
In addition, the quality and large number of spectra allowed us to apply a spectral disentangling technique based on shift-and-add to detect the spectrum of the secondary component, something achieved by the professional teams only recently, and to analyse the star as a SB2 system. Next, we derived the fundamental atmospheric parameters $T_{\mathrm{eff}}$ , log $g$, $\xi$, and also $\nu$ sin $i$ values of each component star.

This campaign is among the first organized within STAROS project (after a running-in phase on the SS433 microquasar). It was a success in terms of the number of participants, the enthusiasm and the quality of the measurements. Our initial objective was to achieve a radial velocity accuracy of 1 km s$^-1$ (equivalent to a margin of error of 0.022 Å). Not only was this objective achieved, but it was also surpassed, offering excellent prospects for future campaigns involving similar targets (i.e. spectroscopic binaries). The radial velocity curve obtained stands up to comparison with the most recent professional publications. Our advantage in this context is first and foremost sheer numbers. Indeed, the very essence of this campaign lay in the unprecedented coordination of numerous spectrographs operating simultaneously over a given period of time, with the aim of achieving radial velocity accuracy through averaging. This innovative approach, rarely used by professionals, was crowned with success, thanks to the flexibility of amateur in organizing this type of operation.

\section*{Acknowledgments}

We acknowledge Jaroslav Merc for useful advices and for kindly review this manuscript before submission.

\section*{Conflict of interest}

The authors declare no potential conflict of interests.

\nocite{*}
\bibliography{Wiley-ASNA}

\begin{thebibliography}{}

\bibitem [\protect \citeauthoryear {%
{Astropy Collaboration}%
\ \protect \BOthers {.}}{%
{Astropy Collaboration}%
\ \protect \BOthers {.}}{%
{\protect \APACyear {2013}}%
}]{%
astropy2013}
\APACinsertmetastar {%
astropy2013}%
\begin{APACrefauthors}%
{Astropy Collaboration}%
, Robitaille, T\BPBI P.%
, Tollerud, E\BPBI J.%
\ et al.\end{APACrefauthors}%
\unskip\
\newblock
\APACrefYearMonthDay{2013}{}{},
\newblock
\unskip
\newblock
\APACjournalVolNumPages{\aap}{558}{}{A33}.
\newblock
\begin{APACrefDOI} \doi{10.1051/0004-6361/201322068} \end{APACrefDOI}
\PrintBackRefs{\CurrentBib}

\bibitem [\protect \citeauthoryear {%
Bedding%
, Hey%
\BCBL {}\ \BBA {} Murphy%
}{%
Bedding%
\ \protect \BOthers {.}}{%
{\protect \APACyear {2019}}%
}]{%
Bedding2019}
\APACinsertmetastar {%
Bedding2019}%
\begin{APACrefauthors}%
Bedding, T\BPBI R.%
, Hey, D\BPBI R.%
\BCBL {}\ \BBA {} Murphy, S\BPBI J.%
\end{APACrefauthors}%
\unskip\
\newblock
\APACrefYearMonthDay{2019}{}{},
\newblock
\unskip
\newblock
\APACjournalVolNumPages{Research Notes of the American Astronomical Society}{3}{}{163}.
\PrintBackRefs{\CurrentBib}

\bibitem [\protect \citeauthoryear {%
Bischoff%
\ \protect \BOthers {.}}{%
Bischoff%
\ \protect \BOthers {.}}{%
{\protect \APACyear {2017}}%
}]{%
bischoff2017}
\APACinsertmetastar {%
bischoff2017}%
\begin{APACrefauthors}%
Bischoff, R.%
, Mugrauer, M.%
, Zehe, T.%
\ et al.\end{APACrefauthors}%
\unskip\
\newblock
\APACrefYearMonthDay{2017}{}{},
\newblock
\unskip
\newblock
\APACjournalVolNumPages{Astronomische Nachrichten}{338}{6}{671--679}.
\PrintBackRefs{\CurrentBib}

\bibitem [\protect \citeauthoryear {%
Castelli%
\ \BBA {} Kurucz%
}{%
Castelli%
\ \BBA {} Kurucz%
}{%
{\protect \APACyear {2004}}%
}]{%
castelli2004new}
\APACinsertmetastar {%
castelli2004new}%
\begin{APACrefauthors}%
Castelli, F.%
\BCBT {}\ \BBA {} Kurucz, R\BPBI L.%
\end{APACrefauthors}%
\unskip\
\newblock
\APACrefYearMonthDay{2004}{}{},
\newblock
\APACrefbtitle {New Grids of ATLAS9 Model Atmospheres.} {New Grids of ATLAS9 Model Atmospheres.}
\PrintBackRefs{\CurrentBib}

\bibitem [\protect \citeauthoryear {%
Gray%
}{%
Gray%
}{%
{\protect \APACyear {1999}}%
}]{%
Gray1982}
\APACinsertmetastar {%
Gray1982}%
\begin{APACrefauthors}%
Gray, R\BPBI O.%
\end{APACrefauthors}%
\unskip\
\newblock
\APACrefYearMonthDay{1999}{}{},
\newblock
\unskip
\newblock
\APACjournalVolNumPages{Astrophysics Source Code Library, ascl:9910.002}{}{}{}.
\PrintBackRefs{\CurrentBib}

\bibitem [\protect \citeauthoryear {%
Harper%
}{%
Harper%
}{%
{\protect \APACyear {1907}}%
}]{%
harper1907}
\APACinsertmetastar {%
harper1907}%
\begin{APACrefauthors}%
Harper, W\BPBI E.%
\end{APACrefauthors}%
\unskip\
\newblock
\APACrefYearMonthDay{1907}{}{},
\newblock
\unskip
\newblock
\APACjournalVolNumPages{Journal of the Royal Astronomical Society of Canada}{1}{}{237}.
\PrintBackRefs{\CurrentBib}

\bibitem [\protect \citeauthoryear {%
Hey%
\ \protect \BOthers {.}}{%
Hey%
\ \protect \BOthers {.}}{%
{\protect \APACyear {2022}}%
}]{%
Hey2022}
\APACinsertmetastar {%
Hey2022}%
\begin{APACrefauthors}%
Hey, D\BPBI R.%
, Kochoska, A.%
, Monier, R.%
\ et al.\end{APACrefauthors}%
\unskip\
\newblock
\APACrefYearMonthDay{2022}{}{},
\newblock
\unskip
\newblock
\APACjournalVolNumPages{Monthly Notices of the Royal Astronomical Society}{511}{}{2648--2658}.
\PrintBackRefs{\CurrentBib}

\bibitem [\protect \citeauthoryear {%
Hutter%
\ \protect \BOthers {.}}{%
Hutter%
\ \protect \BOthers {.}}{%
{\protect \APACyear {2016}}%
}]{%
Hutter_2016}
\APACinsertmetastar {%
Hutter_2016}%
\begin{APACrefauthors}%
Hutter, D\BPBI J.%
, Zavala, R\BPBI T.%
, Tycner, C.%
\ et al.\end{APACrefauthors}%
\unskip\
\newblock
\APACrefYearMonthDay{2016}{nov}{},
\newblock
\unskip
\newblock
\APACjournalVolNumPages{The Astrophysical Journal Supplement Series}{227}{1}{4}.
\PrintBackRefs{\CurrentBib}

\bibitem [\protect \citeauthoryear {%
Kallinger%
, Iliev%
, Lehmann%
\BCBL {}\ \BBA {} Weiss%
}{%
Kallinger%
\ \protect \BOthers {.}}{%
{\protect \APACyear {2004}}%
}]{%
kallinger2004}
\APACinsertmetastar {%
kallinger2004}%
\begin{APACrefauthors}%
Kallinger, T.%
, Iliev, I.%
, Lehmann, H.%
\BCBL {}\ \BBA {} Weiss, W\BPBI W.%
\end{APACrefauthors}%
\unskip\
\newblock
\APACrefYearMonthDay{2004}{}{},
\newblock
\unskip
\newblock
\APACjournalVolNumPages{The A-Star Puzzle, held in Poprad, Slovakia. Edited by J. Zverko, J. Ziznovsky, S.~J. Adelman, and W.~W. Weiss. IAU Symposium}{224}{}{848--852}.
\PrintBackRefs{\CurrentBib}

\bibitem [\protect \citeauthoryear {%
Murphy%
\ \protect \BOthers {.}}{%
Murphy%
\ \protect \BOthers {.}}{%
{\protect \APACyear {2007}}%
}]{%
murphy2007}
\APACinsertmetastar {%
murphy2007}%
\begin{APACrefauthors}%
Murphy, M\BPBI T.%
, Udem, T.%
, Holzwarth, R.%
\ et al.\end{APACrefauthors}%
\unskip\
\newblock
\APACrefYearMonthDay{2007}{}{},
\newblock
\unskip
\newblock
\APACjournalVolNumPages{Monthly Notices of the Royal Astronomical Society}{380}{2}{839--847}.
\PrintBackRefs{\CurrentBib}

\bibitem [\protect \citeauthoryear {%
Pavlovski%
\ \protect \BOthers {.}}{%
Pavlovski%
\ \protect \BOthers {.}}{%
{\protect \APACyear {2022}}%
}]{%
Pavlovski2022}
\APACinsertmetastar {%
Pavlovski2022}%
\begin{APACrefauthors}%
Pavlovski, K.%
, Hummel, C\BPBI A.%
, Tkachenko, A.%
\ et al.\end{APACrefauthors}%
\unskip\
\newblock
\APACrefYearMonthDay{2022}{}{},
\newblock
\unskip
\newblock
\APACjournalVolNumPages{A\&A}{663}{}{A92}.
\PrintBackRefs{\CurrentBib}

\bibitem [\protect \citeauthoryear {%
Quintero%
, Eenens%
\BCBL {}\ \BBA {} Rauw%
}{%
Quintero%
\ \protect \BOthers {.}}{%
{\protect \APACyear {2020}}%
}]{%
Quintero2020}
\APACinsertmetastar {%
Quintero2020}%
\begin{APACrefauthors}%
Quintero, E\BPBI A.%
, Eenens, P.%
\BCBL {}\ \BBA {} Rauw, G.%
\end{APACrefauthors}%
\unskip\
\newblock
\APACrefYearMonthDay{2020}{}{},
\newblock
\unskip
\newblock
\APACjournalVolNumPages{Astronomische Nachrichten}{341}{6--7}{}.
\PrintBackRefs{\CurrentBib}

\end{thebibliography}

\newpage

\appendix

\section{Individual radial velocity measurements}
Result of individual radial velocity measurements from all observers
reported on the radial velocity curve calculated from the entire campaign STAROS dataset.

\begin{center}
     \includegraphics[width=\linewidth]{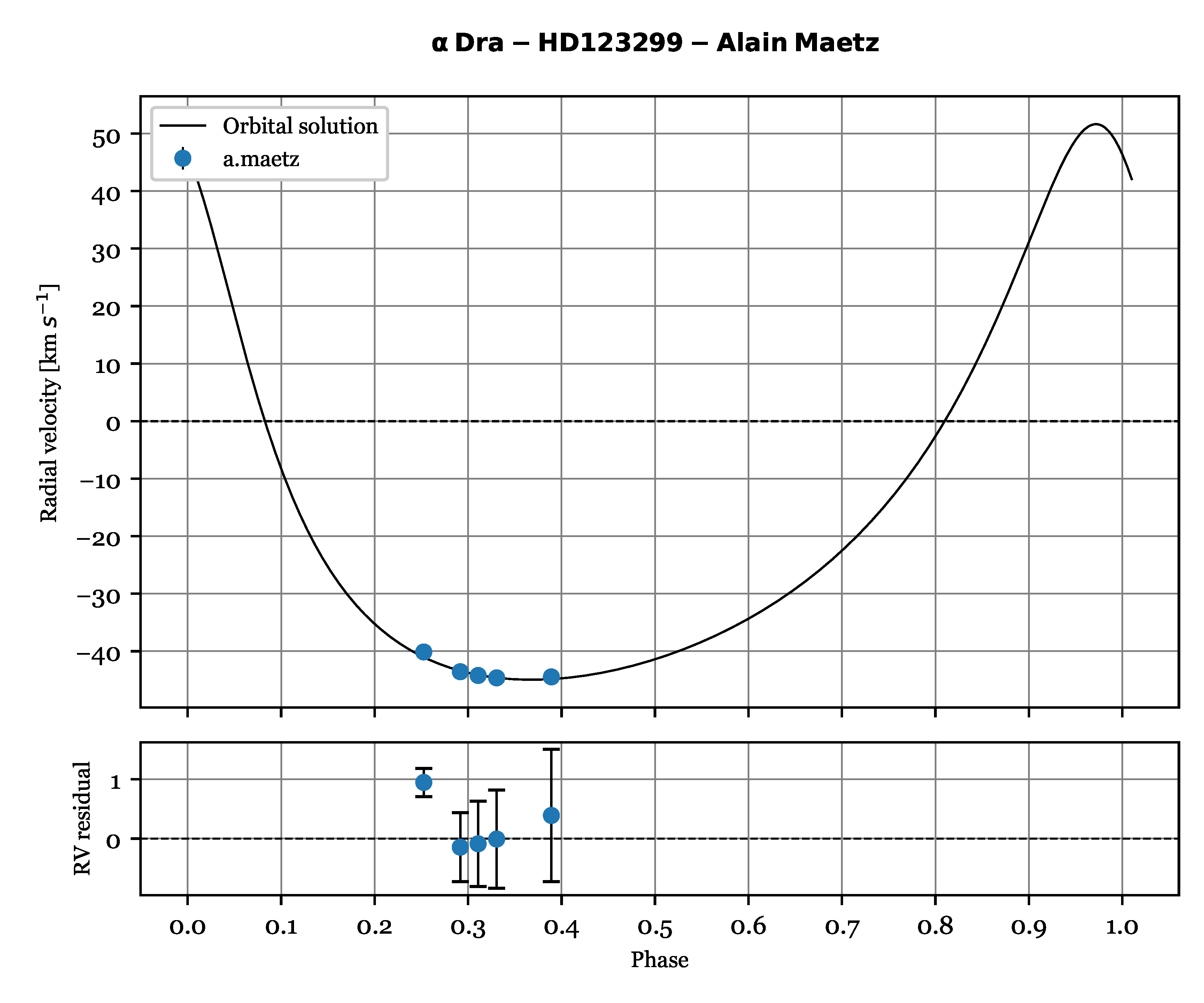}
\end{center}

\begin{center}
     \includegraphics[width=\linewidth]{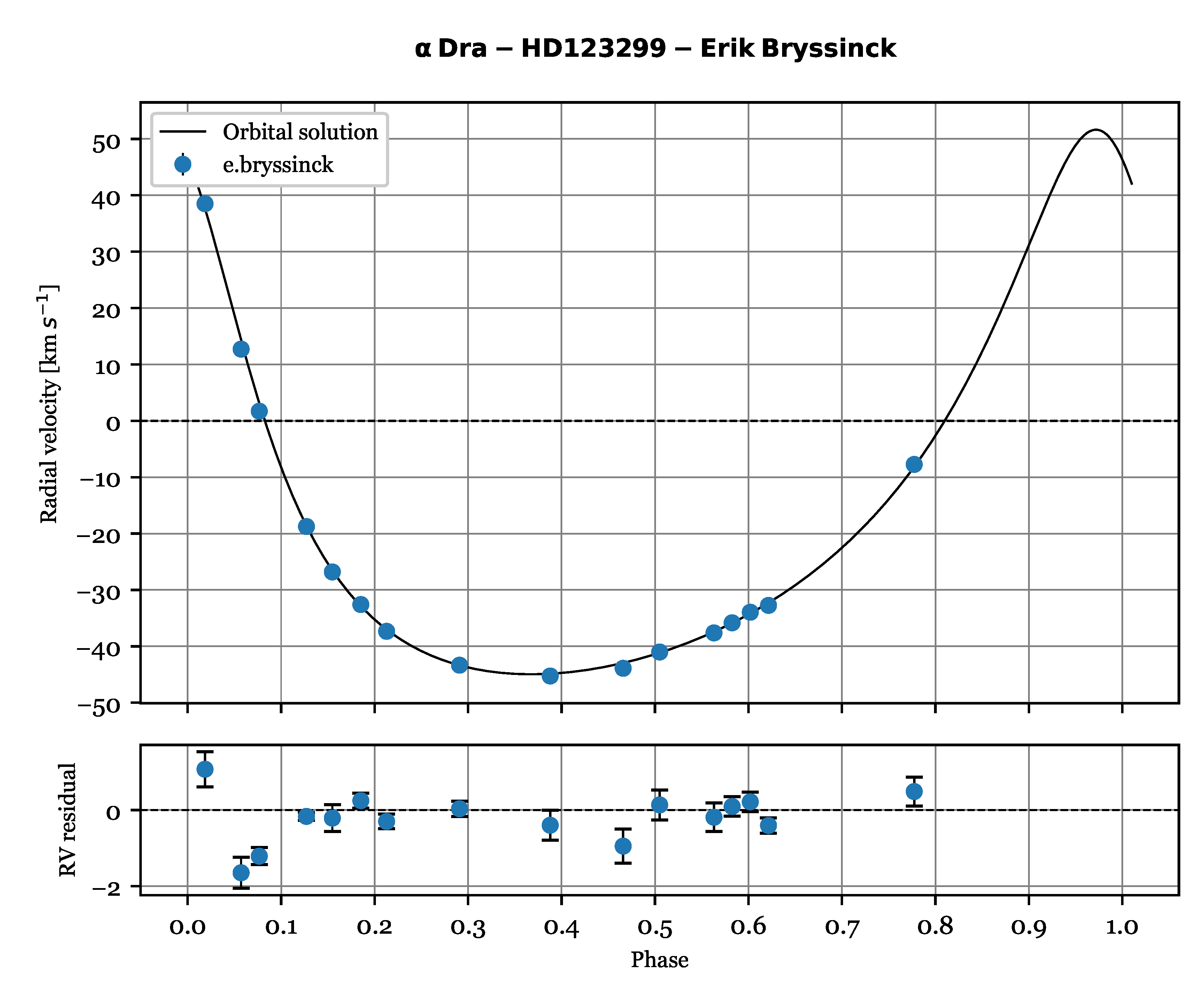}
\end{center}

\begin{center}
     \includegraphics[width=\linewidth]{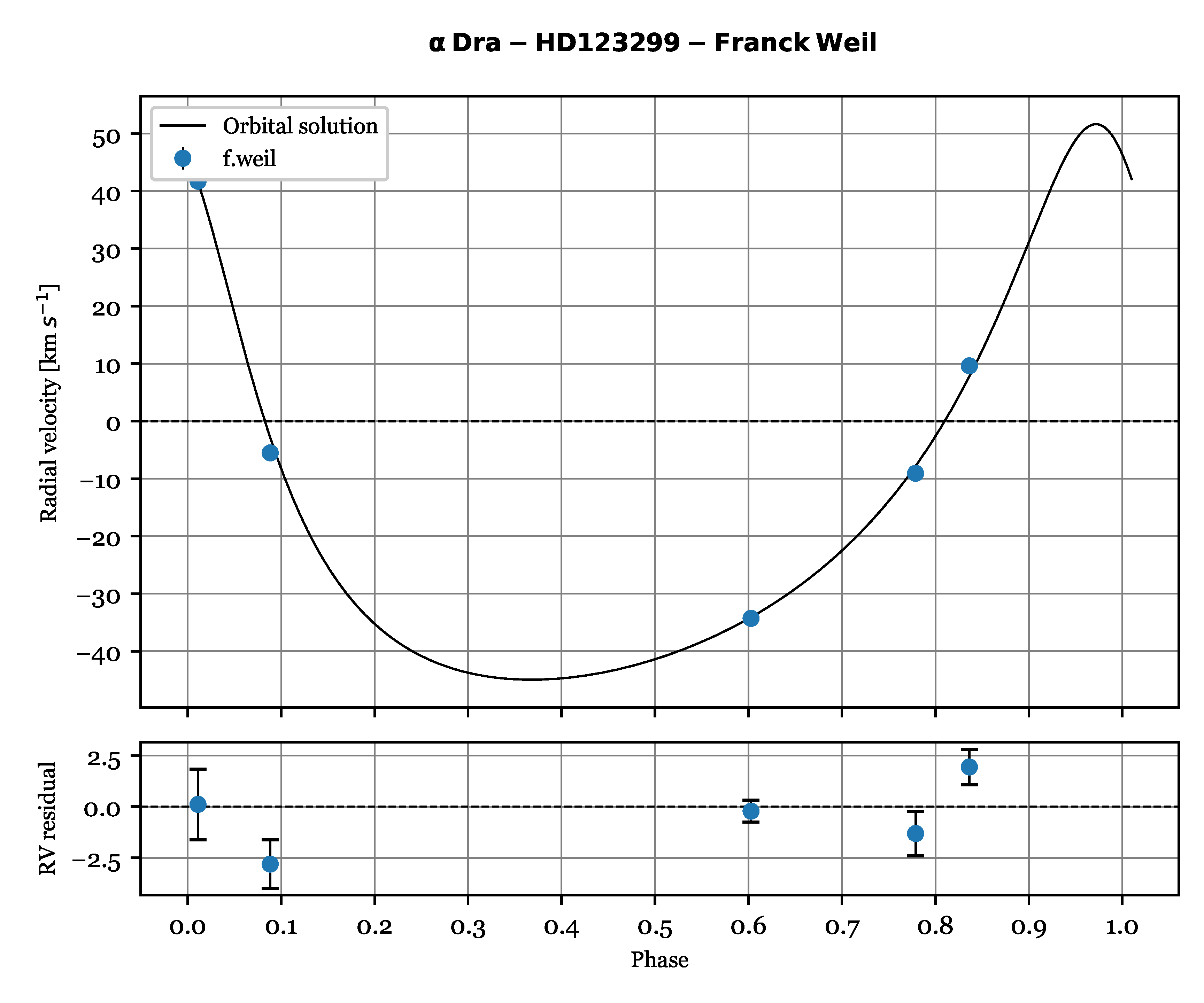}
\end{center}

\begin{center}
     \includegraphics[width=\linewidth]{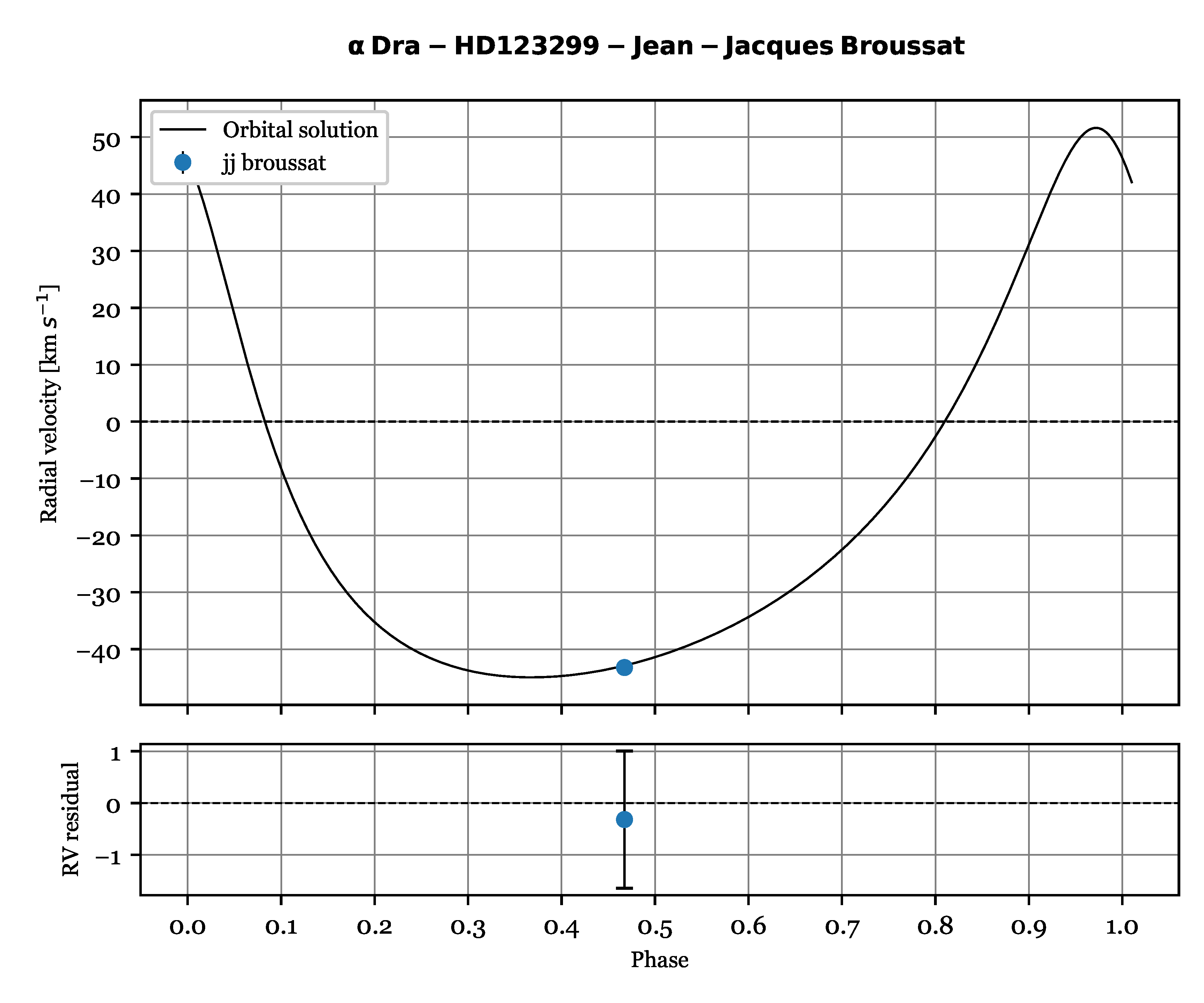}
\end{center}

\begin{center}
     \includegraphics[width=\linewidth]{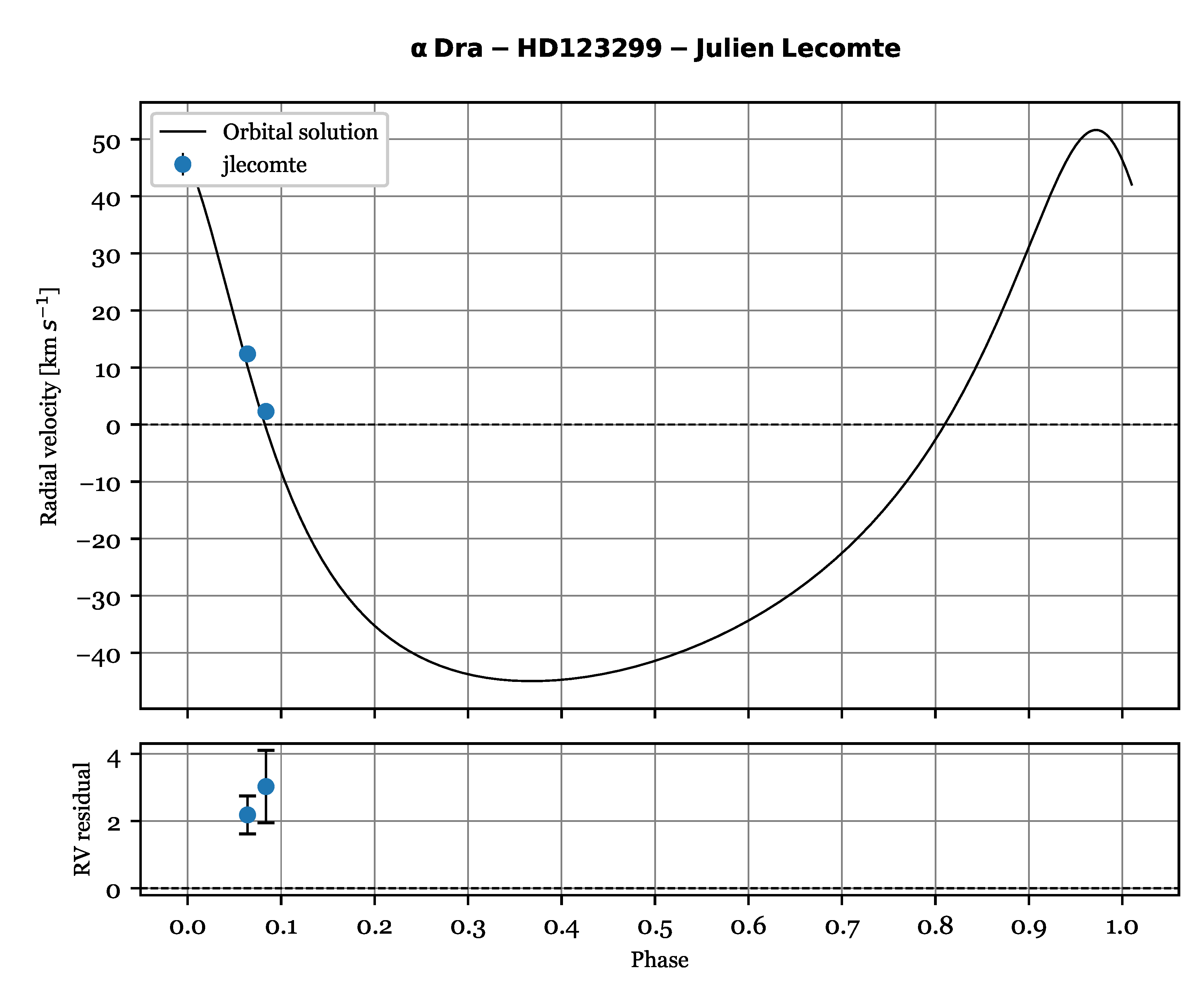}
\end{center}

\begin{center}
     \includegraphics[width=\linewidth]{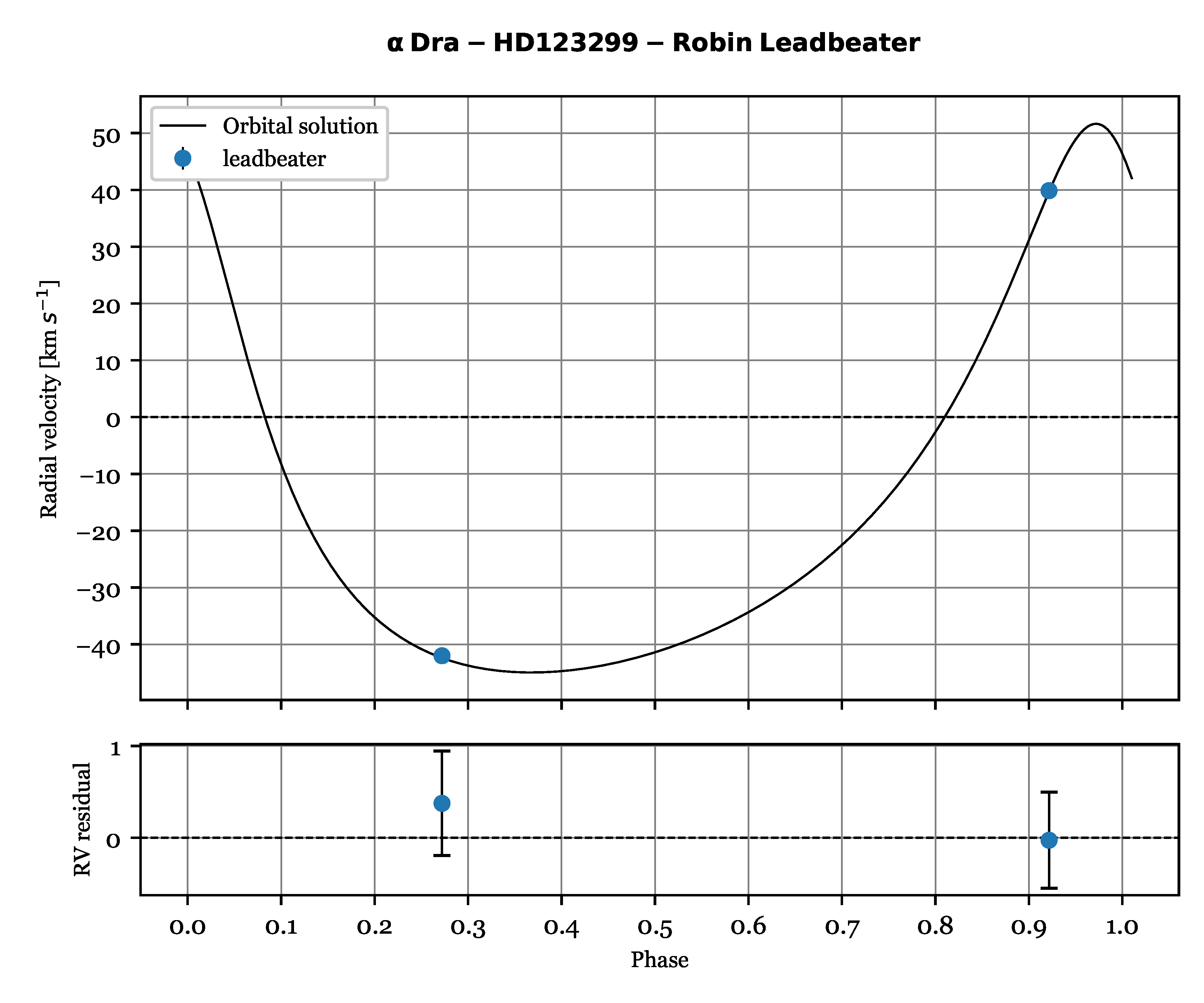}
\end{center}

\begin{center}
     \includegraphics[width=\linewidth]{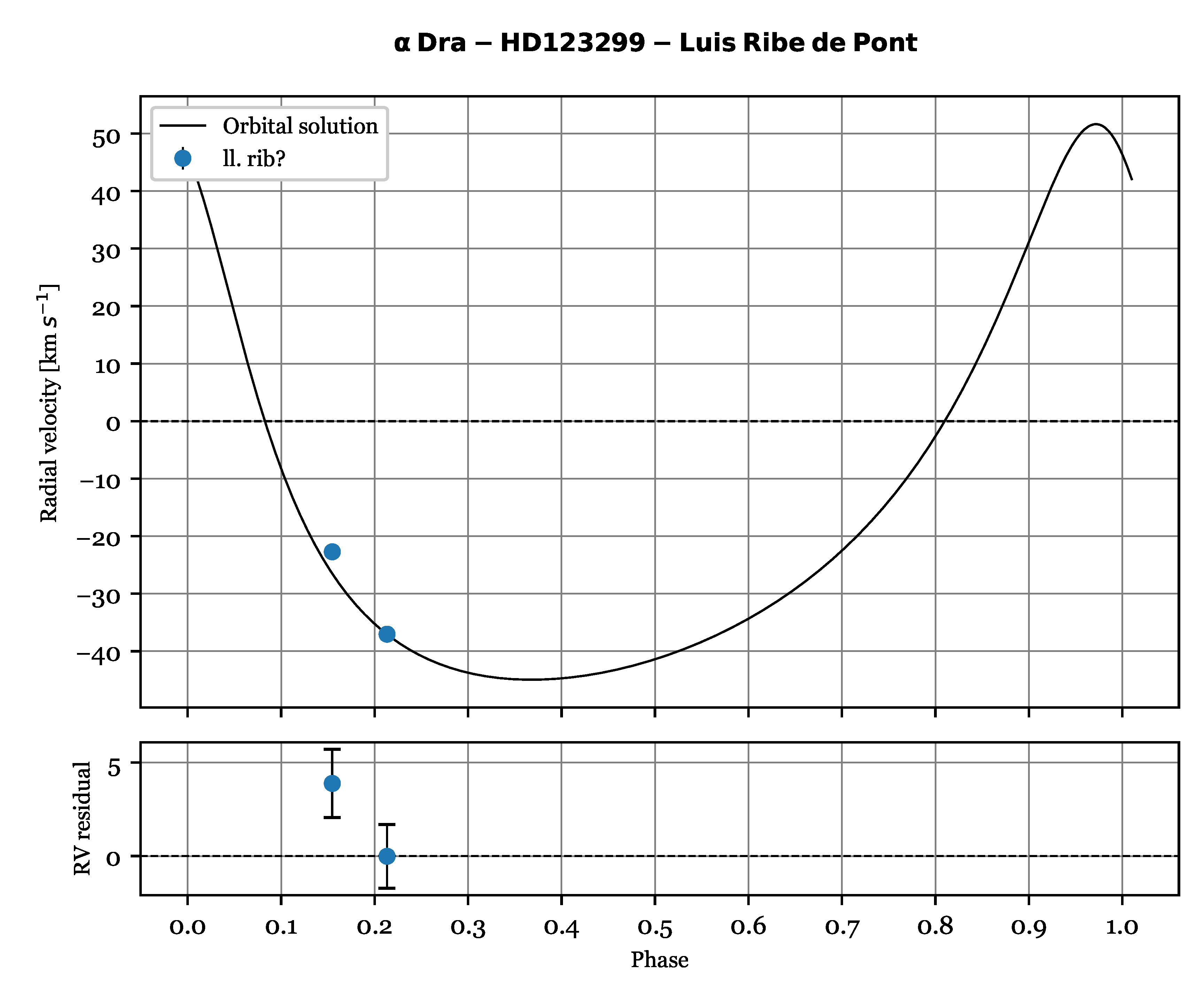}
\end{center}

\begin{center}
     \includegraphics[width=\linewidth]{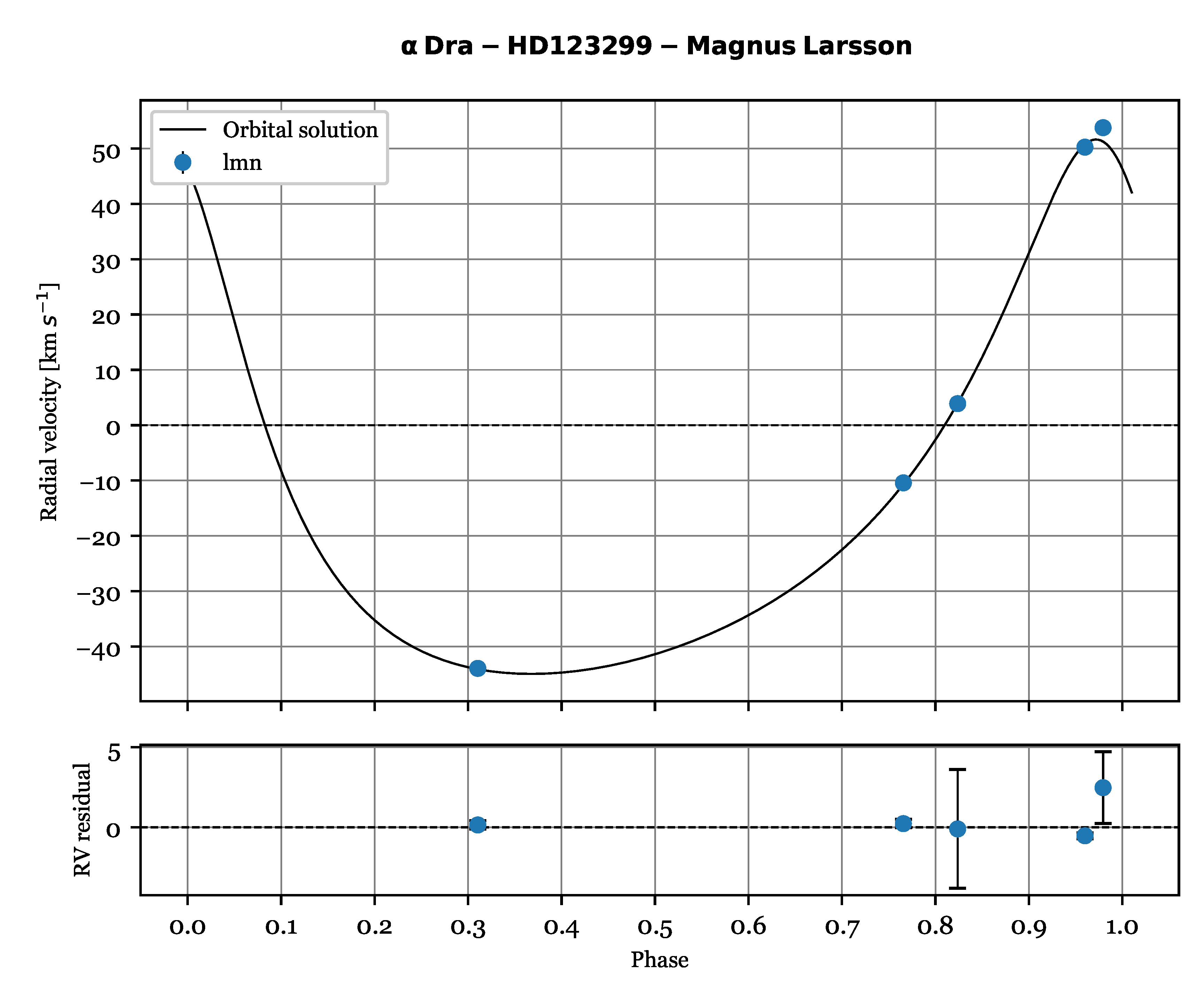}
\end{center}

\begin{center}
     \includegraphics[width=\linewidth]{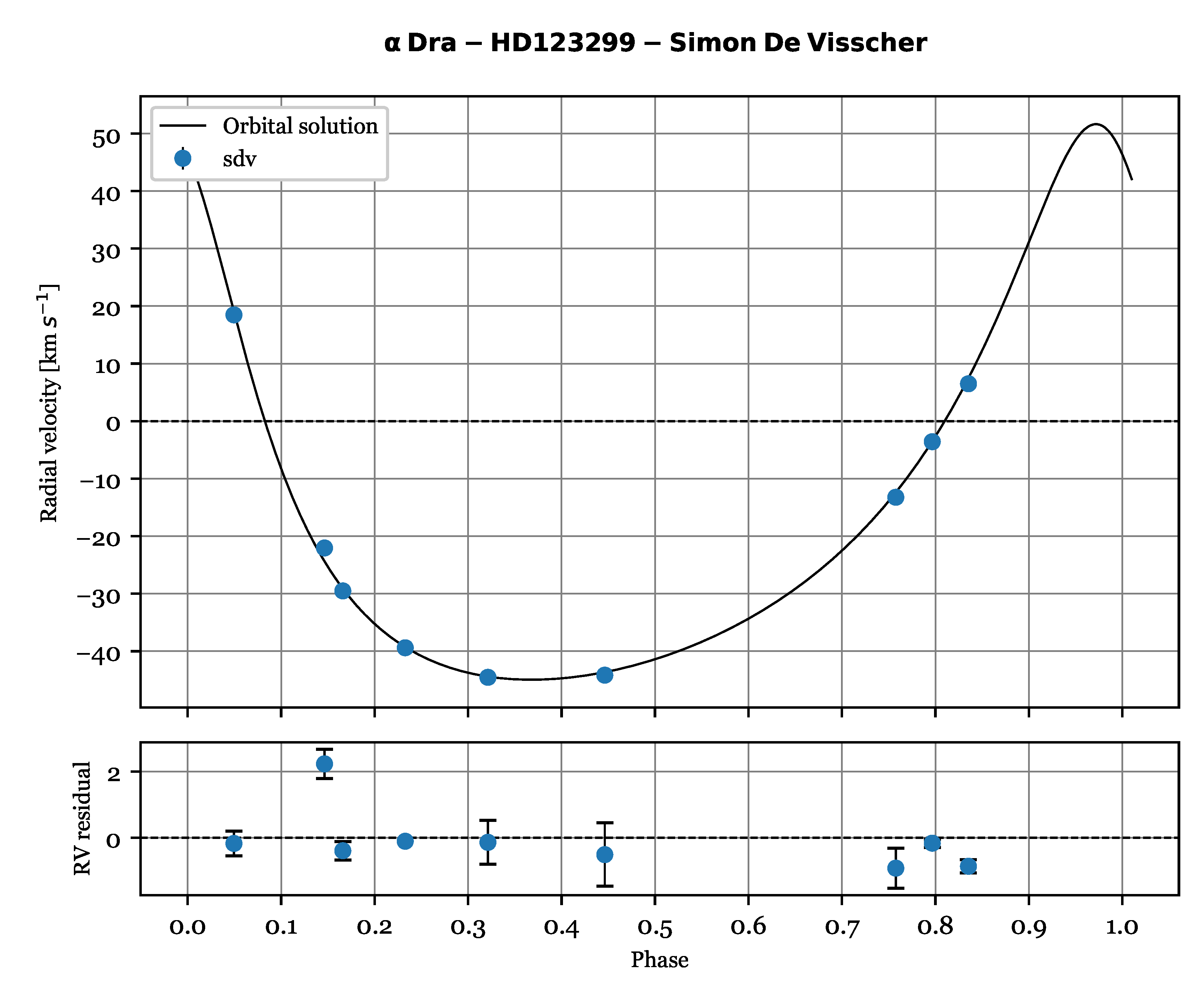}
\end{center}

\begin{center}
     \includegraphics[width=\linewidth]{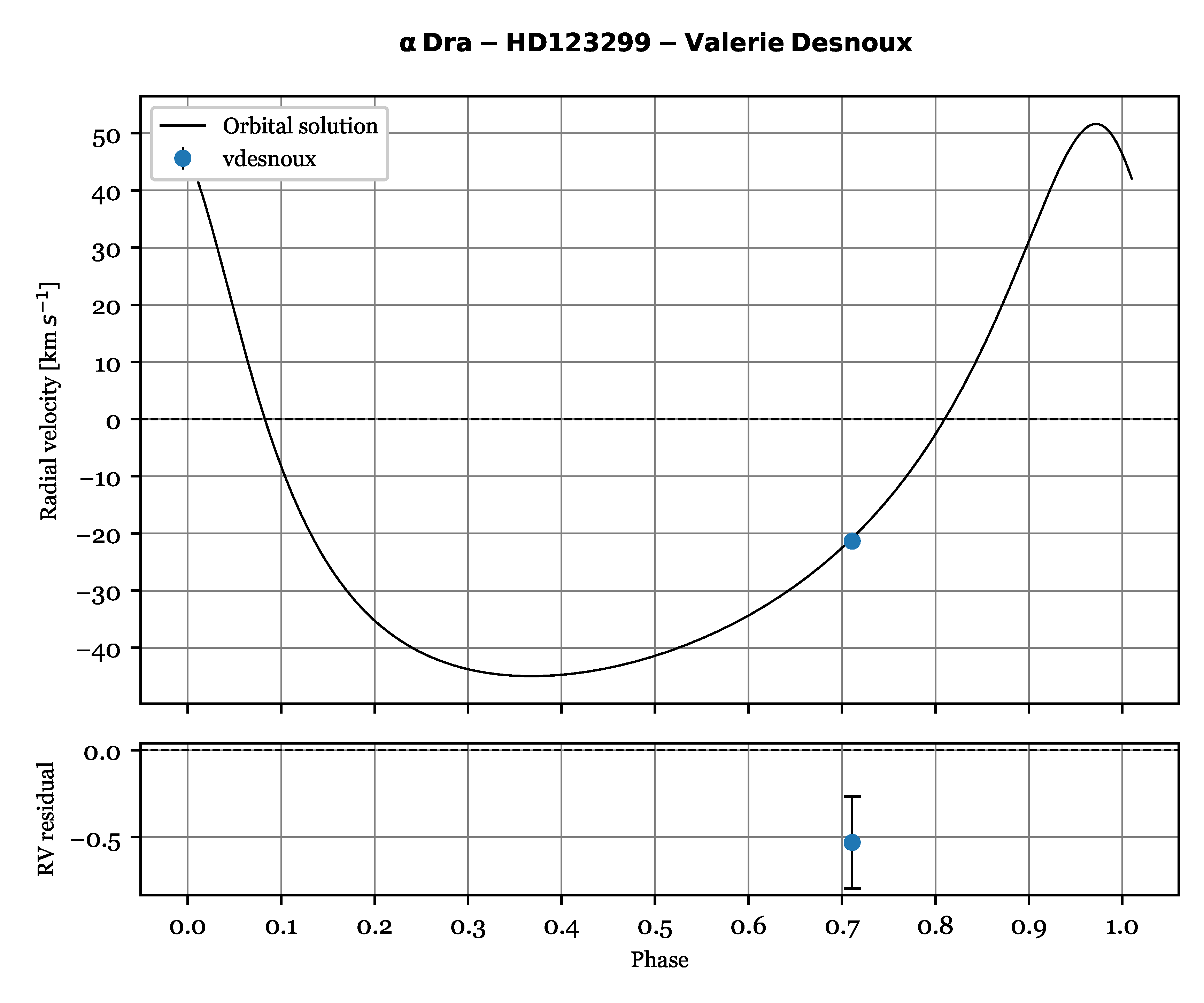}
\end{center}

\begin{center}
     \includegraphics[width=\linewidth]{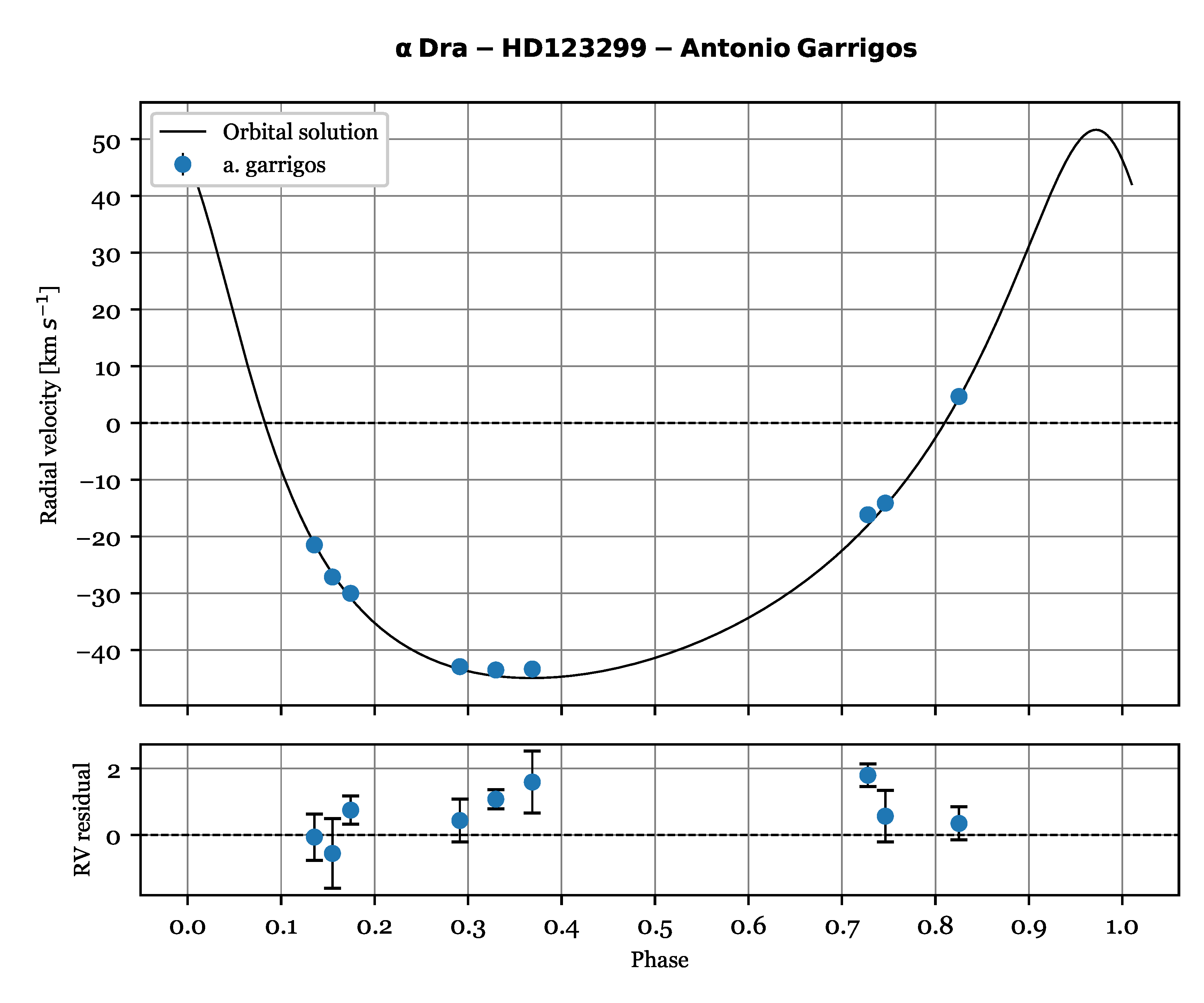}
\end{center}

\begin{center}
     \includegraphics[width=\linewidth]{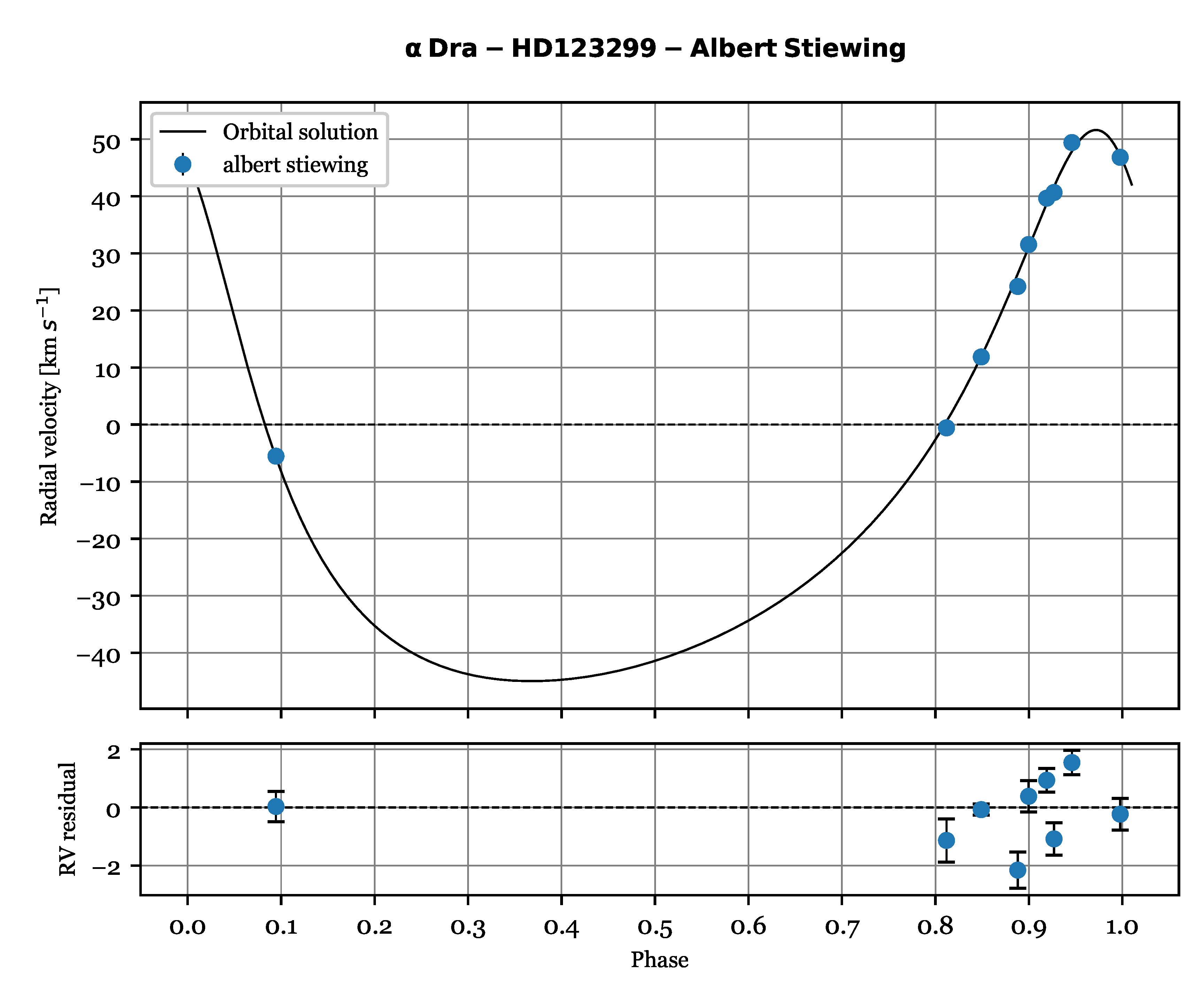}
\end{center}

\begin{center}
     \includegraphics[width=\linewidth]{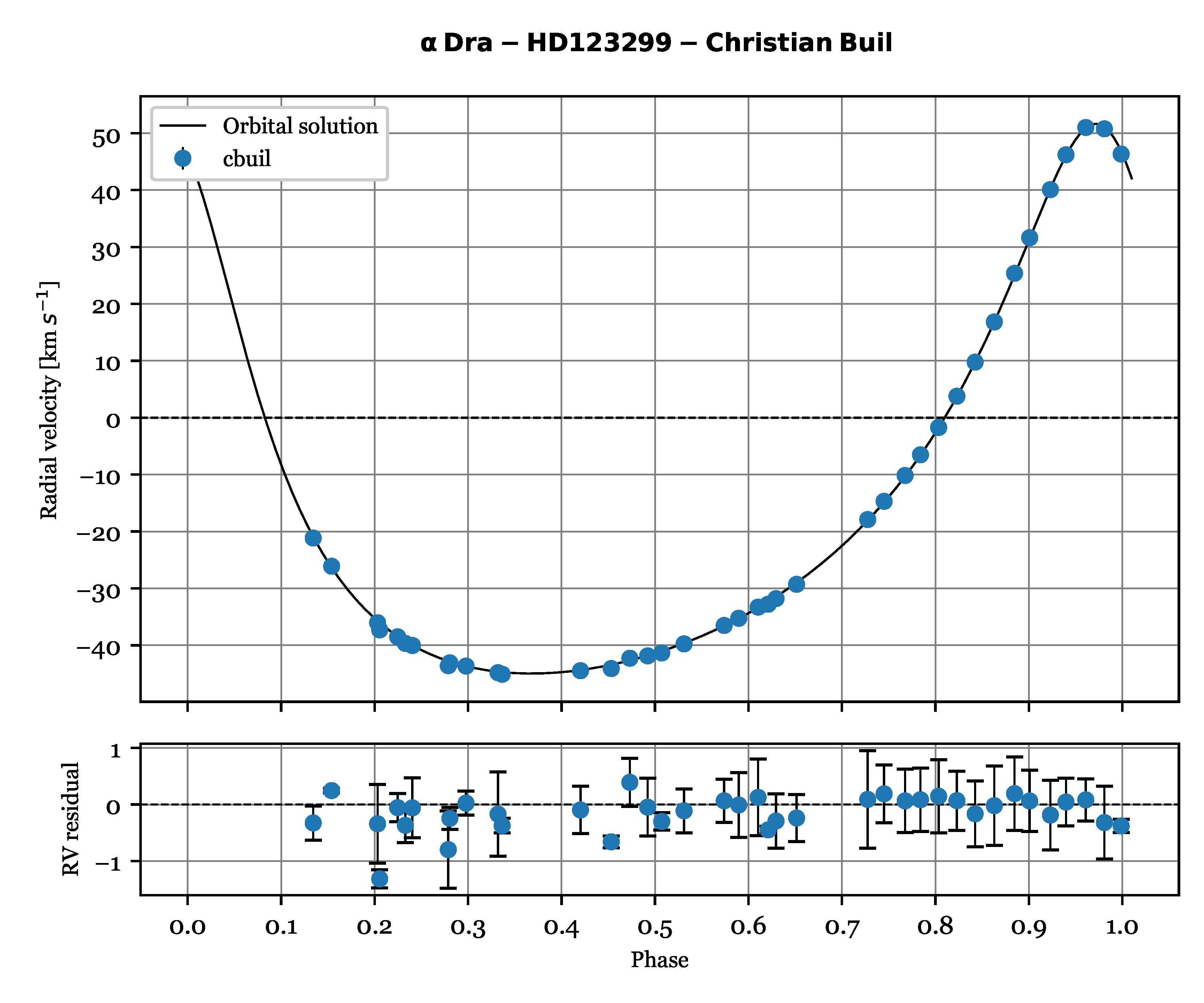}
\end{center}

\begin{center}
     \includegraphics[width=\linewidth]{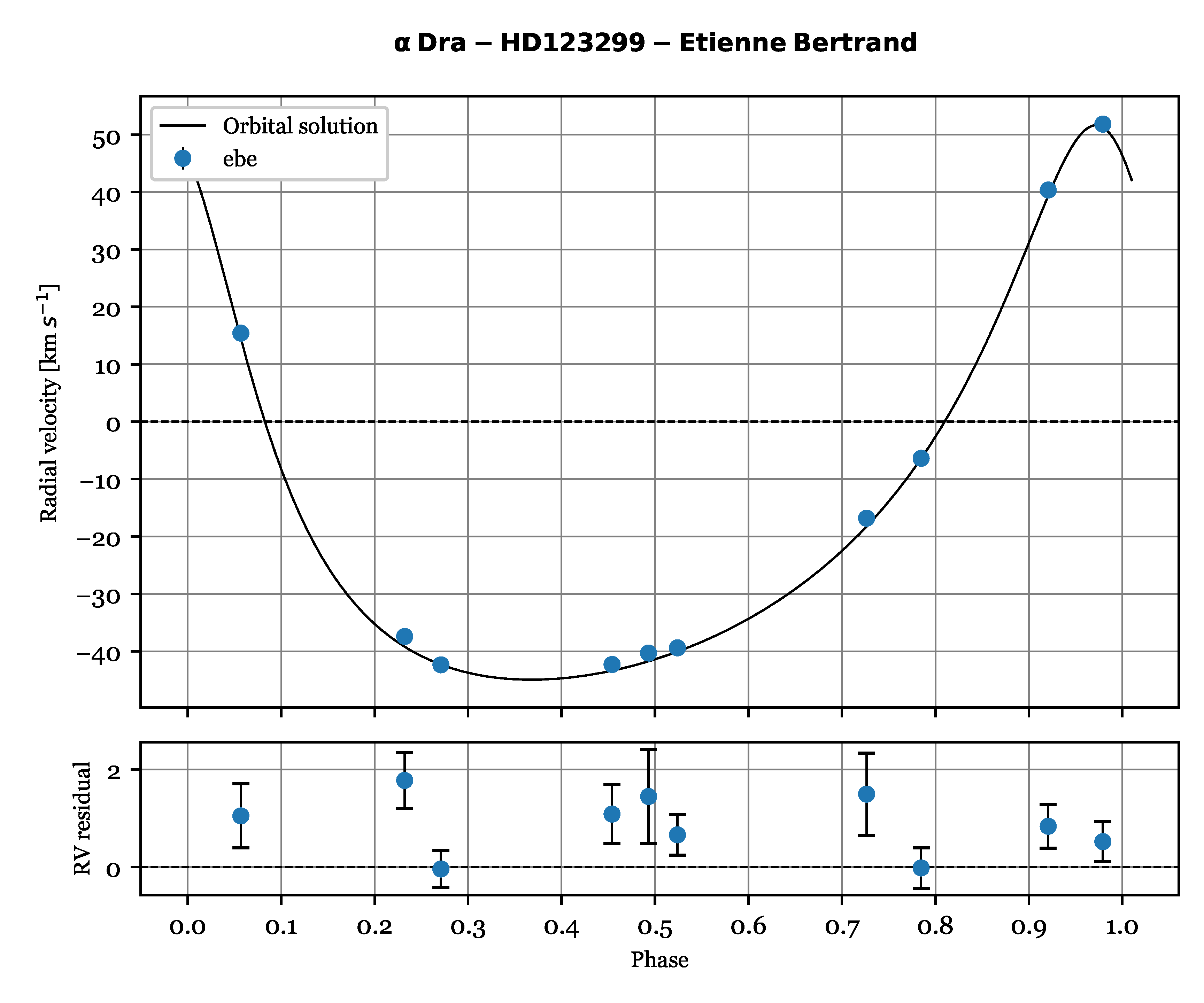}
\end{center}

\begin{center}
     \includegraphics[width=\linewidth]{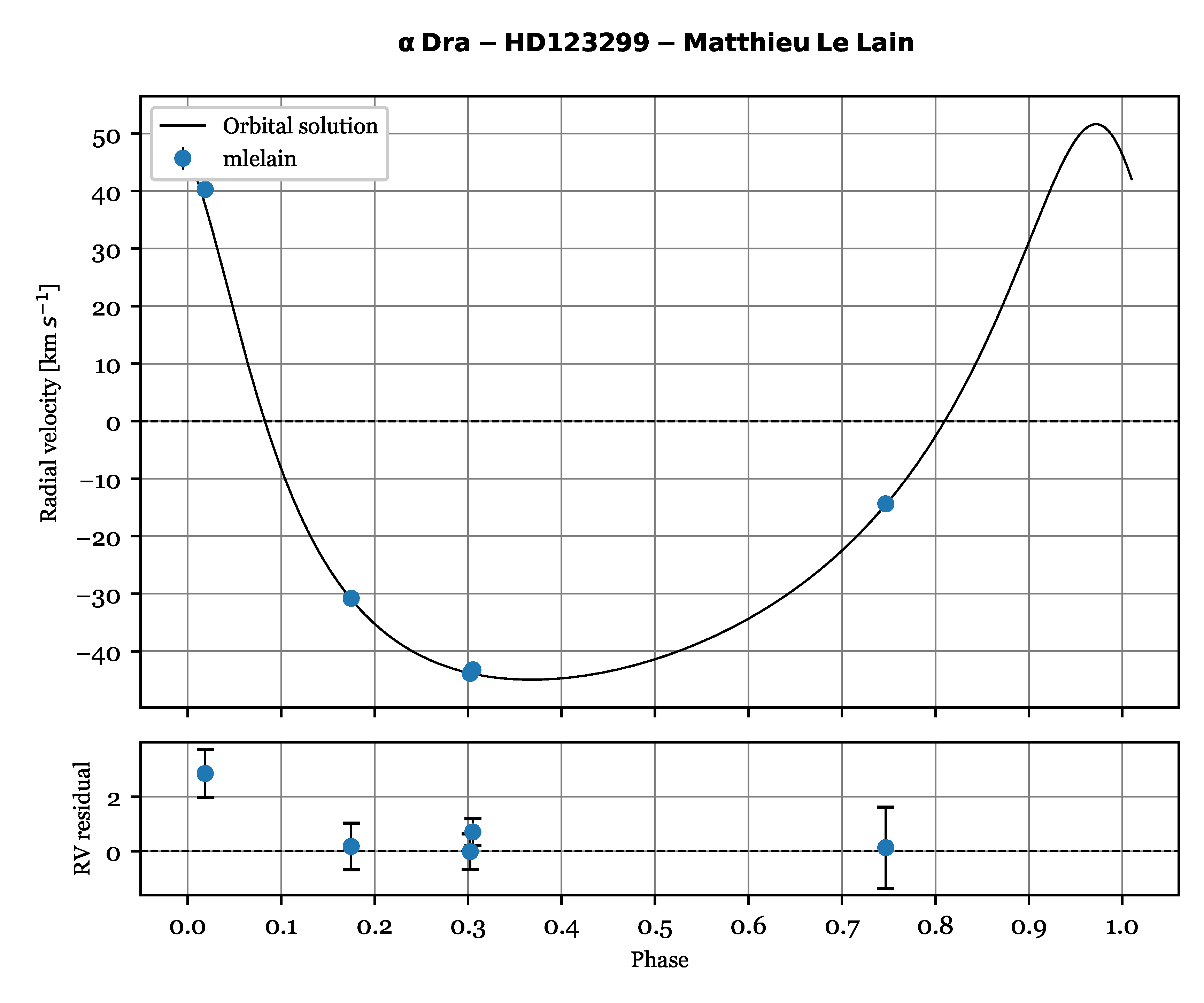}
\end{center}

\begin{center}
     \includegraphics[width=\linewidth]{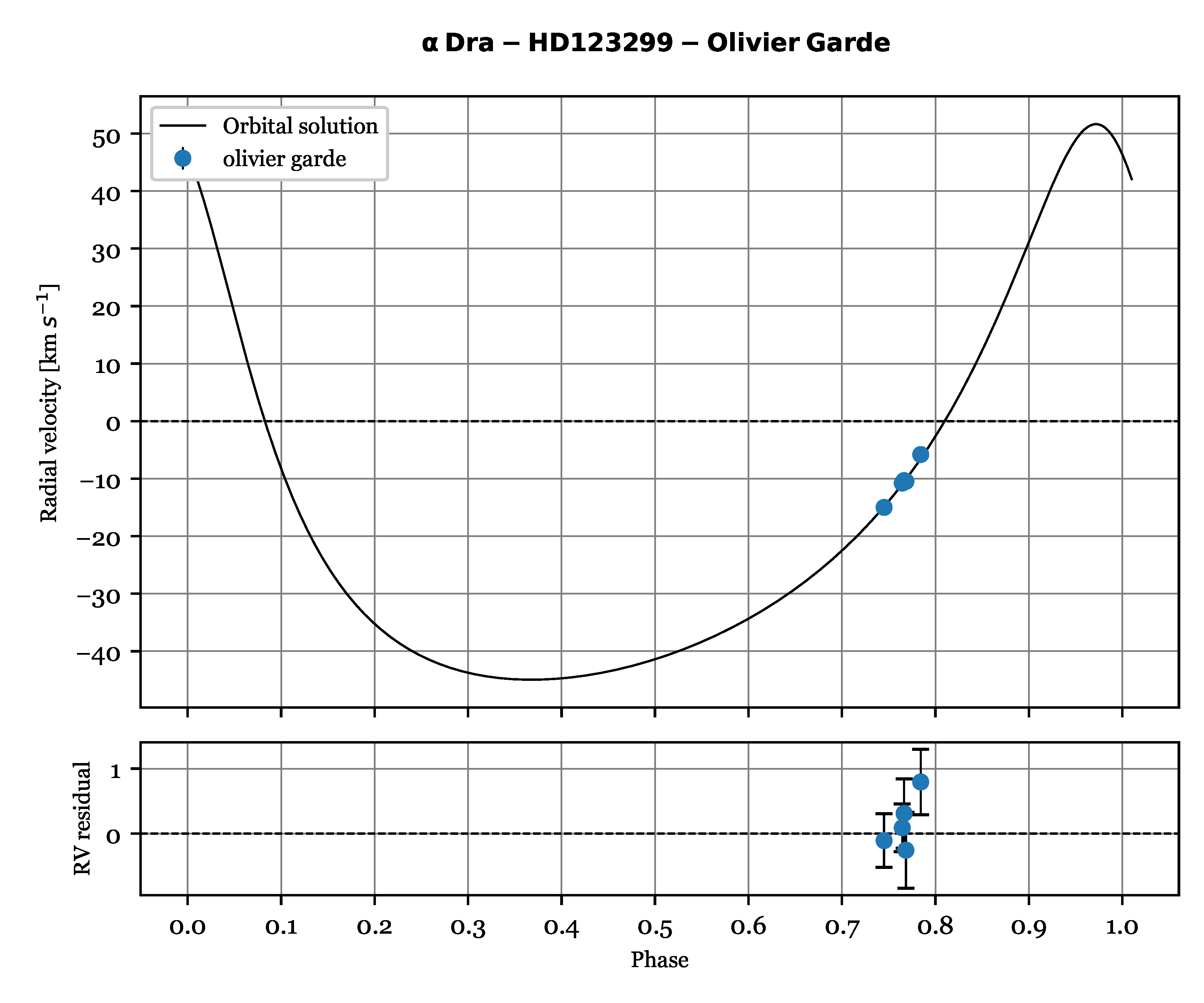}
\end{center}

\begin{center}
     \includegraphics[width=\linewidth]{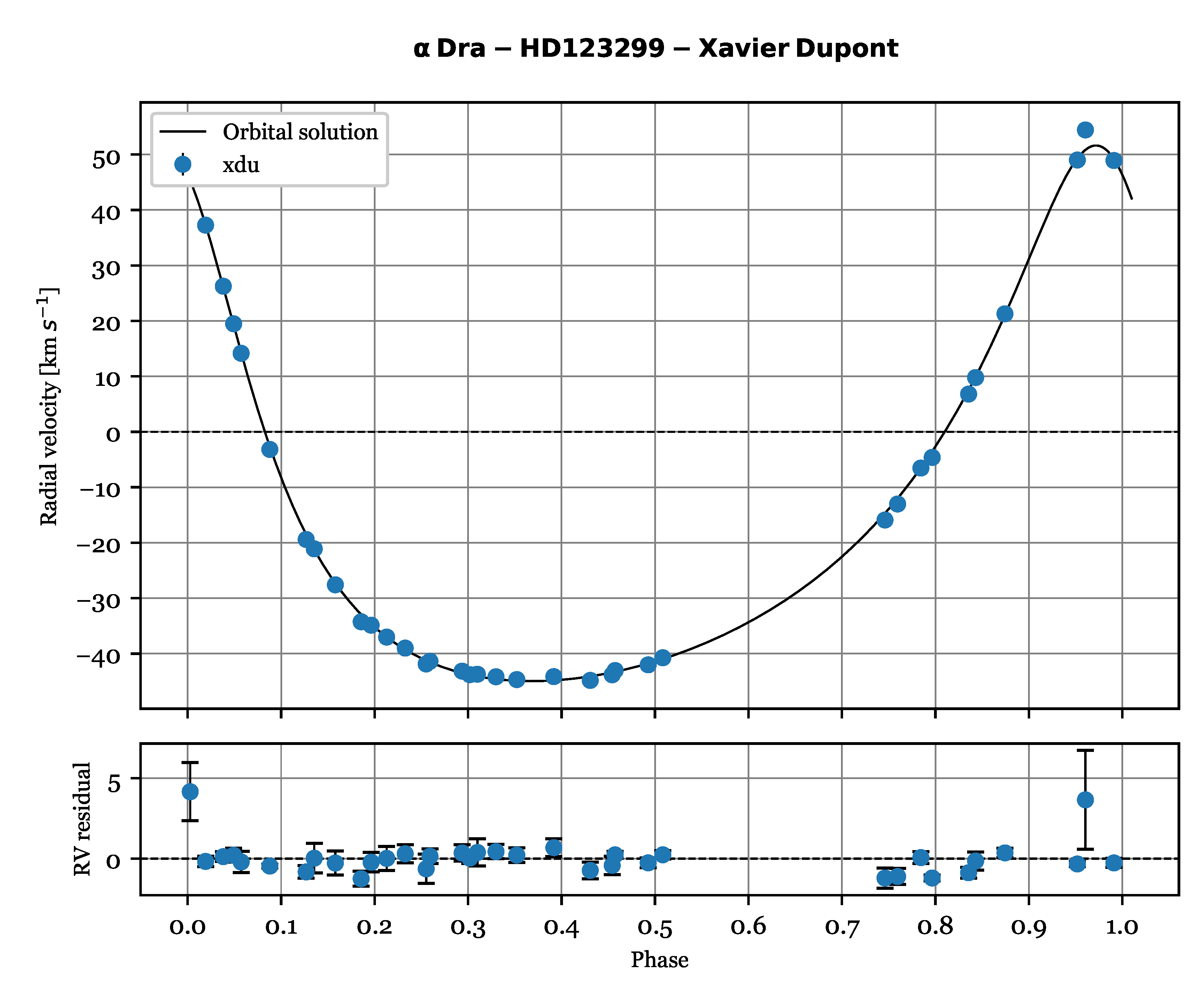}
\end{center}

\begin{center}
     \includegraphics[width=\linewidth]{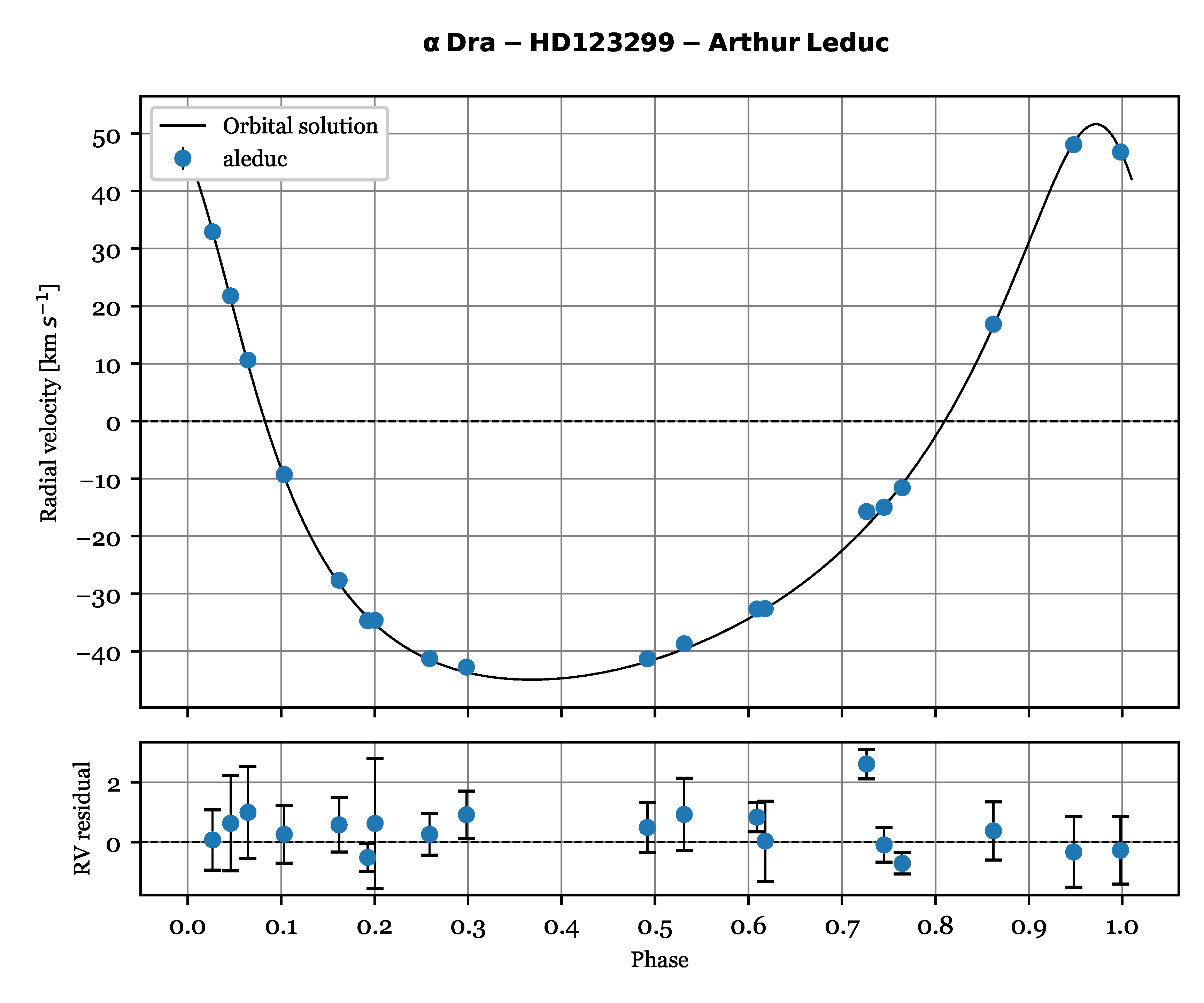}
\end{center}

\begin{center}
     \includegraphics[width=\linewidth]{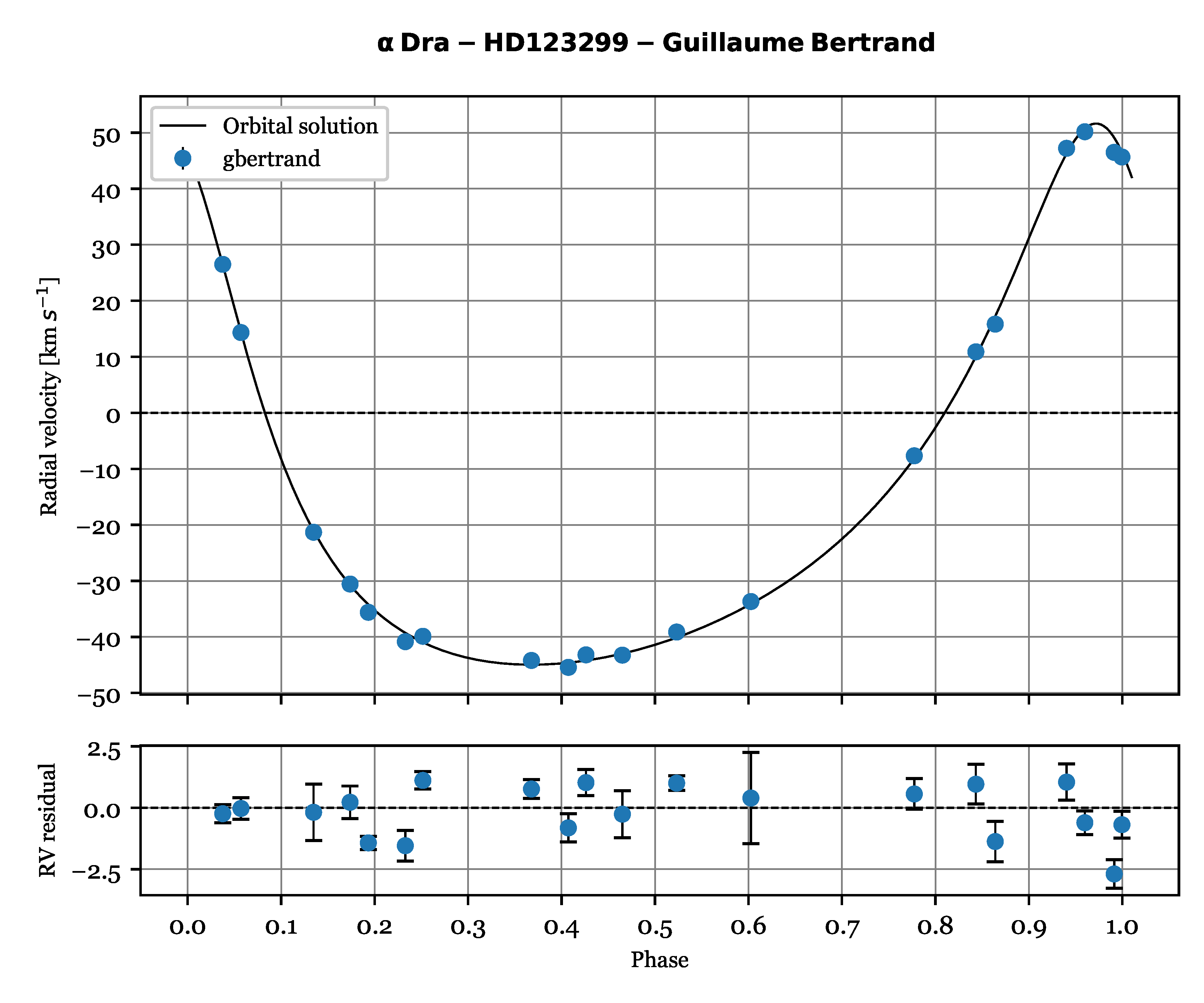}
\end{center}

\begin{center}
     \includegraphics[width=\linewidth]{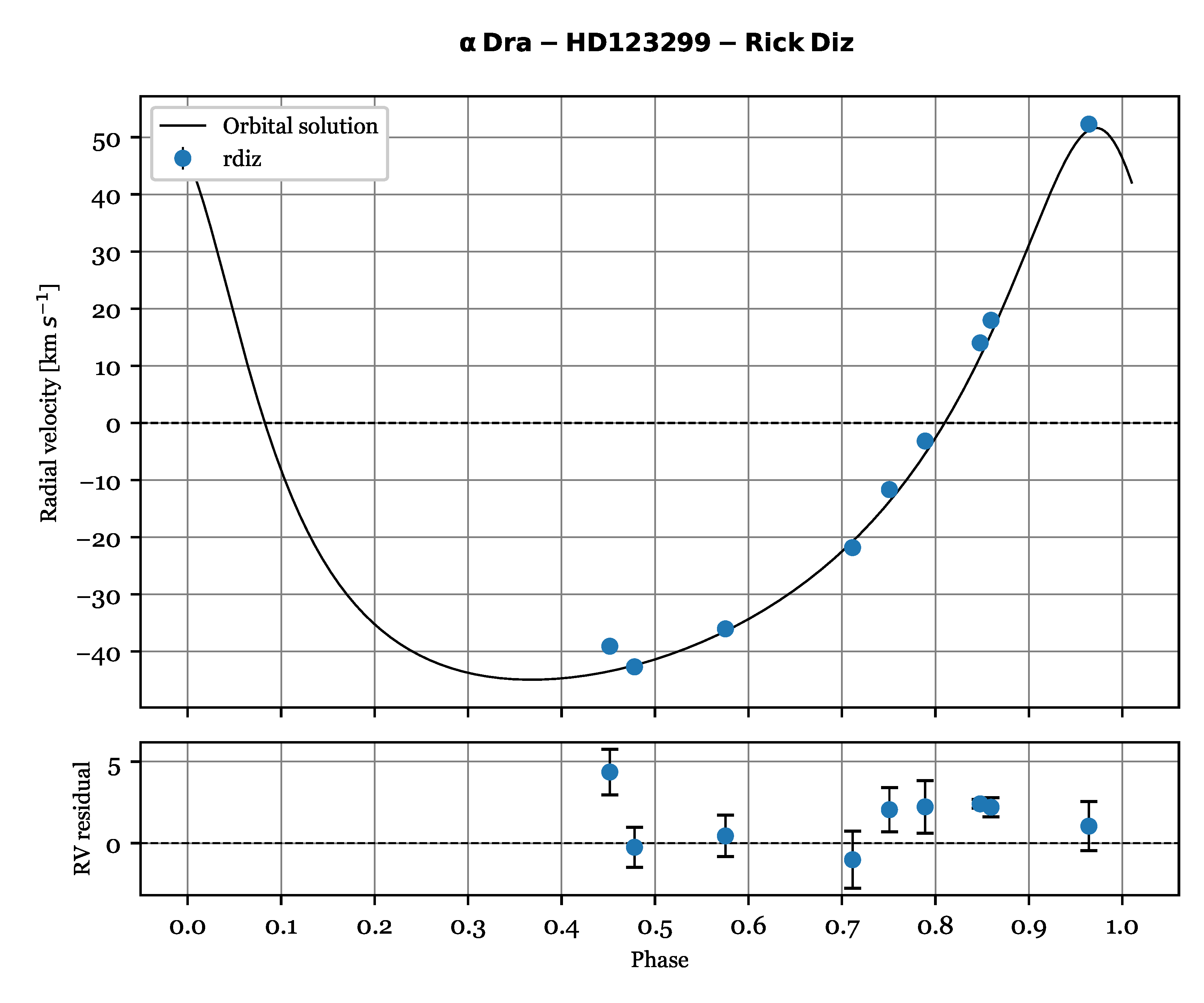}
\end{center}

\begin{center}
     \includegraphics[width=\linewidth]{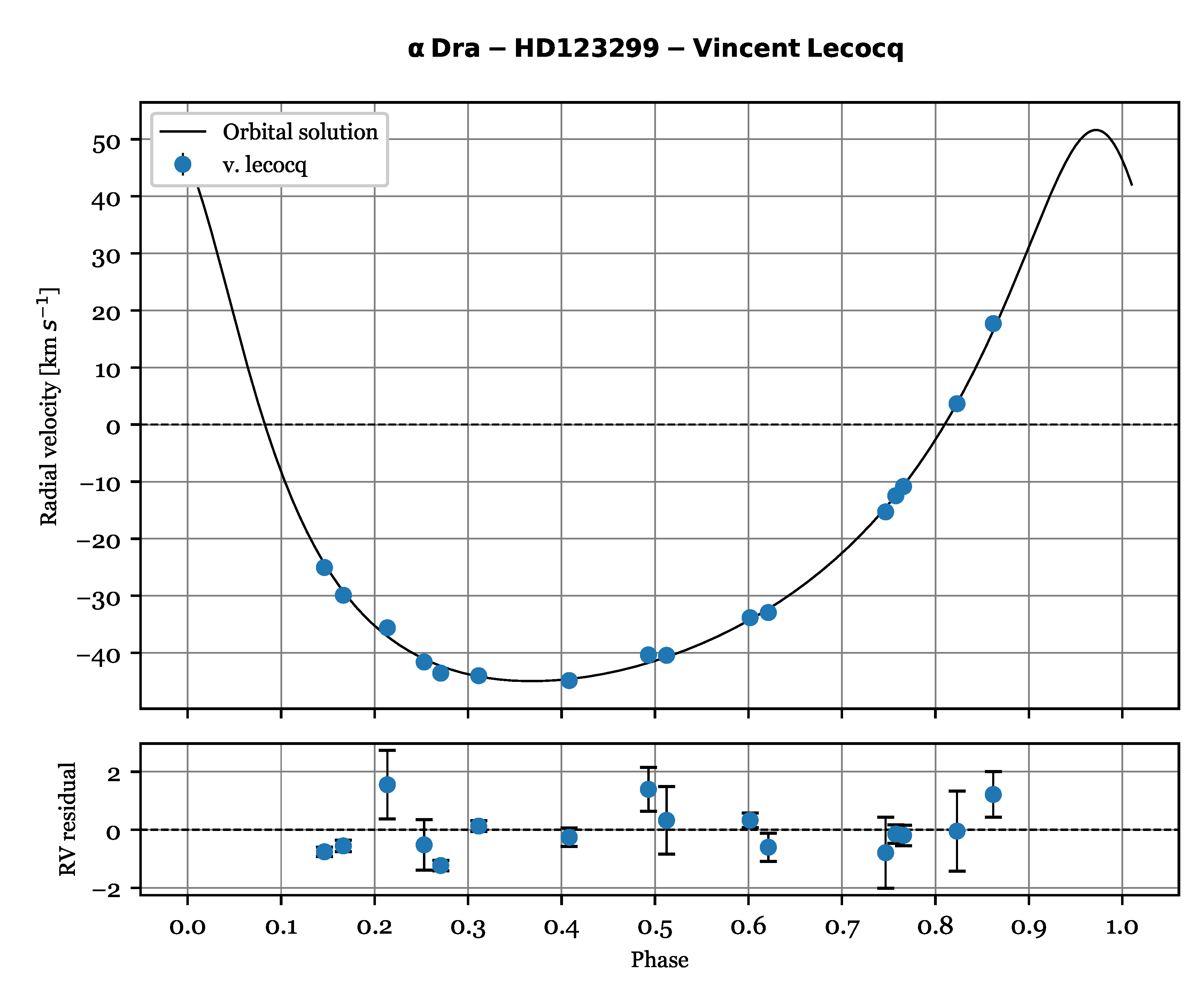}
\end{center}

\begin{center}
\begin{table*}[t]%
\centering
   \caption{RV measurments of $\alpha$ Dra.\label{table:measurments_vr}}%
\tabcolsep=0pt%
\begin{tabular*}{500pt}{@{\extracolsep\fill}lccc}
\toprule
    \textbf{Date of obs. (JD)} &\textbf{Time exposure (s)} & \textbf{Phase} & \textbf{RV (km s$^{-1}$)} \\
\midrule
    2459713.479 & 2000.00 & 0.8640065698 & 15.83310112\\
    2459727.399 & 2400.00 & 0.1346988324 & -21.31081673\\
    2459733.412 & 2400.00 & 0.2516376052 & -39.88529497\\
    2459739.385 & 2000.00 & 0.3678104206 & -44.1961094\\
    2459742.384 & 1800.00 & 0.4261249267 & -43.19685\\
    2459744.388 & 1200.00 & 0.4650972493 & -43.24588345\\
    2459747.372 & 1200.00 & 0.5231389215 & -39.12407464\\
    2459874.284 & 1200.00 & 0.991263426 & 46.49556502\\
    2460009.347 & 3600.00 & 0.6179166405 & -32.5984928\\
    2460026.318 & 2700.00 & 0.947962783 & 48.0733925\\
    2460030.364 & 900.00 & 0.02664282048 & 32.91486178\\
    2460031.358 & 900.00 & 0.04597058483 & 21.76193887\\
    2460032.314 & 900.00 & 0.06456534572 & 10.61070709\\
    2460034.306 & 1800.00 & 0.1033111882 & -9.293624391\\
    2460037.322 & 900.00 & 0.161956595 & -27.66382544\\
    2460039.303 & 900.00 & 0.2004860519 & -34.62339715\\
    2460039.441 & 6000.00 & 0.203174874 & -36.02633311\\
    2460039.557 & 3000.00 & 0.2054211188 & -37.27992734\\
    2460040.548 & 3000.00 & 0.224685959 & -38.51304063\\
    2460041.356 & 3000.00 & 0.2404155844 & -40.0314043\\
    2460042.304 & 2700.00 & 0.2588413396 & -41.26513444\\
    2460042.322 & 4200.00 & 0.2592047543 & -41.37400447\\
    2460043.331 & 4200.00 & 0.2788258588 & -43.57953235\\
    2460043.416 & 4800.00 & 0.2804761479 & -43.08070936\\
    2460044.307 & 4800.00 & 0.2978032887 & -43.64672073\\
    2460044.329 & 1800.00 & 0.2982338886 & -42.75707118\\
    2460046.294 & 9000.00 & 0.3364410014 & -45.08406588\\
    2460050.604 & 4800.00 & 0.4202616757 & -44.45525904\\
    2460052.297 & 6600.00 & 0.4531909058 & -44.08181454\\
    2460052.341 & 1804.00 & 0.4540316564 & -42.30402044\\
    2460052.349 & 4800.00 & 0.4542042207 & -43.80548153\\
    2460053.319 & 6000.00 & 0.4730549095 & -42.26936328\\
    2460053.575 & 3600.00 & 0.4780473824 & -42.70145414\\
    2460054.284 & 3600.00 & 0.4918366936 & -41.31462378\\
    2460054.3 & 7800.00 & 0.4921288858 & -41.85102834\\
    2460054.33 & 4800.00 & 0.4927171825 & -42.01349221\\
    2460054.341 & 3600.00 & 0.4929306535 & -40.36097464\\
    2460054.349 & 1807.00 & 0.4930897644 & -40.30989716\\
    2460055.338 & 3000.00 & 0.5123309401 & -40.43536253\\
    2460056.303 & 8400.00 & 0.531084166 & -39.74547481\\
    2460056.311 & 5400.00 & 0.5312524843 & -38.70926921\\
    2460058.51 & 6600.00 & 0.5740074656 & -36.50758368\\
    2460058.578 & 1800.00 & 0.5753286856 & -36.05018245\\
    2460059.311 & 7800.00 & 0.589585085 & -35.24919294\\
    2460060.314 & 1800.00 & 0.6090974969 & -32.67835564\\
    2460060.376 & 5400.00 & 0.6102981912 & -33.2888399\\
    2460061.361 & 7200.00 & 0.6294618338 & -31.79283208\\
    2460062.493 & 6000.00 & 0.6514672194 & -29.28712043\\
\bottomrule
\end{tabular*}
\end{table*}
\end{center}

\begin{center}
\begin{table*}[t]%
\centering
   \caption{RV measurments of $\alpha$ Dra.\label{table:measurments_vr_2}}%
\tabcolsep=0pt%
\begin{tabular*}{500pt}{@{\extracolsep\fill}lccc}
\toprule
    \textbf{Date of obs. (JD)} &\textbf{Time exposure (s)} & \textbf{Phase} & \textbf{RV (km s$^{-1}$)} \\
\midrule
    2460065.573 & 3600.00 & 0.7113745825 & -21.83262302\\
    2460066.342 & 1804.00 & 0.7263177009 & -16.83624905\\
    2460066.344 & 3601.00 & 0.7263620431 & -15.72103175\\
    2460066.406 & 4800.00 & 0.7275646988 & -17.8946271\\
    2460066.412 & 1508.00 & 0.7276934365 & -16.1900943\\
    2460067.305 & 2400.00 & 0.7450538255 & -14.98280746\\
    2460067.306 & 5400.00 & 0.745076664 & -14.69934254\\
    2460067.307 & 6077.00 & 0.7450985635 & -14.99637016\\
    2460067.355 & 4800.00 & 0.7460339761 & -15.89931288\\
    2460067.373 & 1206.00 & 0.7463829136 & -14.13447433\\
    2460067.391 & 3000.00 & 0.7467224736 & -15.29606866\\
    2460067.4 & 4800.00 & 0.7468922284 & -14.3740166\\
    2460067.6 & 1800.00 & 0.7507903547 & -11.67436821\\
    2460068.305 & 7243.00 & 0.7645049448 & -10.77249961\\
    2460068.306 & 2400.00 & 0.7645126625 & -11.57445495\\
    2460068.367 & 3600.00 & 0.7656980229 & -10.42265082\\
    2460068.376 & 3000.00 & 0.7658786111 & -10.8409909\\
    2460068.407 & 8453.00 & 0.7664843655 & -10.33770398\\
    2460068.462 & 6000.00 & 0.7675476719 & -10.15461926\\
    2460068.505 & 9662.00 & 0.768388832 & -10.47371338\\
    2460069.308 & 8400.00 & 0.7840137401 & -6.517393443\\
    2460069.321 & 7245.00 & 0.7842662892 & -5.807090301\\
    2460069.33 & 6000.00 & 0.7844381169 & -6.533738351\\
    2460069.343 & 2417.00 & 0.7846892289 & -6.385210997\\
    2460069.565 & 3600.00 & 0.7890082106 & -3.176349847\\
    2460070.307 & 7800.00 & 0.8034450256 & -1.716410748\\
    2460070.738 & 1206.00 & 0.8118199427 & -0.5705741081\\
    2460071.313 & 7200.00 & 0.8229930894 & 3.775108901\\
    2460071.319 & 3000.00 & 0.8231123787 & 3.660395009\\
    2460071.354 & 3600.00 & 0.823790176 & 3.895692674\\
    2460071.427 & 2931.00 & 0.825224587 & 4.64360278\\
    2460072.316 & 6600.00 & 0.8425020469 & 9.767436828\\
    2460072.341 & 5400.00 & 0.8429938349 & 9.790559092\\
    2460072.355 & 4500.00 & 0.8432733882 & 10.89073667\\
    2460072.592 & 3608.00 & 0.8478801084 & 14.00493856\\
    2460072.653 & 1206.00 & 0.8490690213 & 11.85567512\\
    2460073.313 & 3600.00 & 0.8618902662 & 17.69700964\\
    2460073.324 & 3600.00 & 0.8621067013 & 16.85548872\\
    2460073.369 & 6000.00 & 0.8629889935 & 16.8216162\\
    2460074.483 & 6600.00 & 0.8846479022 & 25.36974959\\
    2460074.66 & 1206.00 & 0.8880884142 & 24.20350036\\
    2460075.315 & 7200.00 & 0.9008253334 & 31.63165064\\
    2460076.338 & 3012.00 & 0.920725447 & 40.34351658\\
    2460076.382 & 820.00 & 0.9215699759 & 39.86352619\\
    2460076.454 & 5400.00 & 0.9229792946 & 40.08277962\\
\bottomrule
\end{tabular*}
\end{table*}
\end{center}

\begin{center}
\begin{table*}[t]%
\centering
   \caption{RV measurments of $\alpha$ Dra.\label{table:measurments_vr_3}}%
\tabcolsep=0pt%
\begin{tabular*}{500pt}{@{\extracolsep\fill}lccc}
\toprule
    \textbf{Date of obs. (JD)} &\textbf{Time exposure (s)} & \textbf{Phase} & \textbf{RV (km s$^{-1}$)} \\
\midrule
    2460076.653 & 1078.00 & 0.9268590853 & 40.67338992\\
    2460077.312 & 6600.00 & 0.9396724964 & 46.20979829\\
    2460077.345 & 5400.00 & 0.9403162097 & 47.2091321\\
    2460077.639 & 1367.00 & 0.9460259732 & 49.42838409\\
    2460078.351 & 4500.00 & 0.9598628525 & 50.17472593\\
    2460078.356 & 7200.00 & 0.9599704417 & 50.2497627\\
    2460078.383 & 5400.00 & 0.9604955347 & 54.43742959\\
    2460078.402 & 6000.00 & 0.9608676912 & 50.99922423\\
    2460078.572 & 3600.00 & 0.9641644358 & 52.29803929\\
    2460079.342 & 3014.00 & 0.9791347898 & 51.81925783\\
    2460079.353 & 1800.00 & 0.9793534293 & 53.77179303\\
    2460079.426 & 7200.00 & 0.9807792248 & 50.76495169\\
    2460080.312 & 3600.00 & 0.997999543 & 46.79486517\\
    2460080.348 & 5400.00 & 0.9987043082 & 46.33633479\\
    2460080.39 & 4000.00 & 0.9995242607 & 45.66230037\\
    2460080.554 & 6300.00 & 0.002716403181 & 49.34262306\\
    2460081.369 & 2391.00 & 0.01856834347 & 38.5057239\\
    2460081.381 & 6600.00 & 0.01879913956 & 40.28923241\\
    2460081.399 & 4500.00 & 0.01914308714 & 37.26488398\\
    2460082.342 & 5500.00 & 0.03749382736 & 26.50641372\\
    2460082.375 & 5400.00 & 0.03813725544 & 26.25324839\\
    2460083.341 & 3016.00 & 0.05691044811 & 15.41215898\\
    2460083.346 & 4800.00 & 0.05701609825 & 14.33569106\\
    2460083.355 & 3018.00 & 0.05718100278 & 12.71435966\\
    2460083.357 & 5400.00 & 0.05722814861 & 14.16178848\\
    2460083.707 & 1200.00 & 0.06403111254 & 12.38220193\\
    2460084.354 & 5416.00 & 0.0766197973 & 1.703768323\\
    2460084.722 & 1200.00 & 0.08377532691 & 2.299712426\\
    2460087.321 & 4200.00 & 0.1343106187 & -21.14804905\\
    2460087.378 & 5400.00 & 0.135433081 & -21.09733681\\
    2460087.384 & 1513.00 & 0.1355511183 & -21.49773154\\
    2460088.334 & 1389.00 & 0.1540205252 & -26.10159194\\
    2460088.369 & 3600.00 & 0.1546907706 & -22.71485406\\
    2460088.374 & 4814.00 & 0.1547985876 & -26.80645825\\
    2460088.379 & 1832.00 & 0.1548877223 & -27.1532061\\
    2460088.539 & 5400.00 & 0.158001629 & -27.58535431\\
    2460089.35 & 3600.00 & 0.173766431 & -30.55944766\\
    2460089.381 & 1513.00 & 0.1743715337 & -30.03763016\\
    2460089.412 & 6300.00 & 0.1749855138 & -30.805353\\
    2460090.318 & 1800.00 & 0.1926043626 & -34.68763044\\
    2460090.353 & 4800.00 & 0.1932789933 & -35.60097062\\
    2460090.507 & 5400.00 & 0.1962790016 & -34.8668533\\
    2460091.359 & 4814.00 & 0.212845539 & -37.33500525\\
    2460091.362 & 6300.00 & 0.2129064689 & -37.01496686\\
    2460091.383 & 3600.00 & 0.2133029173 & -37.04789138\\
    2460091.402 & 3000.00 & 0.2136734352 & -35.60823273\\
    2460092.351 & 3279.00 & 0.232145454 & -37.42335586\\
    2460092.382 & 5400.00 & 0.2327365743 & -38.99420743\\
\bottomrule
\end{tabular*}
\end{table*}
\end{center}

\begin{center}
\begin{table*}[t]%
\centering
   \caption{RV measurments of $\alpha$ Dra.\label{table:measurments_vr_4}}%
\tabcolsep=0pt%
\begin{tabular*}{500pt}{@{\extracolsep\fill}lccc}
\toprule
    \textbf{Date of obs. (JD)} &\textbf{Time exposure (s)} & \textbf{Phase} & \textbf{RV (km s$^{-1}$)} \\
\midrule
    2460092.384 & 3067.00 & 0.2327892047 & -39.40522824\\
    2460092.388 & 3600.00 & 0.2328518097 & -40.8420254\\
    2460092.393 & 5400.00 & 0.2329513414 & -39.66739443\\
    2460092.409 & 3090.00 & 0.2332603131 & -37.76909122\\
    2460093.403 & 2100.00 & 0.2526003198 & -40.13000953\\
    2460093.422 & 2400.00 & 0.2529645847 & -41.59558134\\
    2460093.536 & 4500.00 & 0.2551782326 & -41.85929562\\
    2460094.335 & 3600.00 & 0.270714829 & -43.55414929\\
    2460094.347 & 3016.00 & 0.2709516889 & -42.36372078\\
    2460094.406 & 368.00 & 0.2721104401 & -42.00738124\\
    2460095.371 & 3610.00 & 0.2908624909 & -43.34142866\\
    2460095.389 & 1824.00 & 0.2912145286 & -42.94319737\\
    2460095.412 & 3000.00 & 0.2916707004 & -43.56475045\\
    2460095.513 & 4500.00 & 0.2936326705 & -43.14801355\\
    2460096.353 & 5400.00 & 0.3099744642 & -43.71162892\\
    2460096.374 & 9000.00 & 0.3103733546 & -43.95614296\\
    2460096.394 & 2700.00 & 0.3107627894 & -44.21751715\\
    2460096.425 & 6600.00 & 0.3113634347 & -44.00209767\\
    2460097.366 & 1513.00 & 0.3296603982 & -43.53123675\\
    2460097.377 & 5400.00 & 0.3298853168 & -44.16617311\\
    2460097.413 & 4500.00 & 0.3305780148 & -44.63283894\\
    2460097.489 & 7200.00 & 0.332067446 & -44.8122926\\
    2460098.521 & 4500.00 & 0.3521215027 & -44.67422715\\
    2460099.374 & 2144.00 & 0.3687117536 & -43.36637589\\
    2460100.358 & 4814.00 & 0.3878606345 & -45.26211582\\
    2460100.423 & 6213.00 & 0.3891183227 & -44.46383671\\
    2460100.558 & 4500.00 & 0.3917419465 & -44.13885124\\
    2460101.357 & 3600.00 & 0.4072849418 & -45.42977876\\
    2460101.405 & 6000.00 & 0.4082066173 & -44.85497847\\
    2460102.562 & 2700.00 & 0.4307168024 & -44.82009657\\
    2460103.362 & 4200.00 & 0.4462724779 & -44.16428211\\
    2460103.637 & 1800.00 & 0.4516245872 & -39.09466559\\
    2460104.372 & 3610.00 & 0.4659181026 & -43.89364053\\
    2460104.446 & 3535.00 & 0.4673606893 & -43.22315394\\
    2460106.378 & 3611.00 & 0.5049347942 & -41.00675781\\
    2460106.494 & 6600.00 & 0.5071801863 & -41.33460759\\
    2460106.553 & 3600.00 & 0.5083258638 & -40.73877559\\
    2460107.36 & 3104.00 & 0.5240330307 & -39.4045669\\
    2460109.368 & 3610.00 & 0.5630735814 & -37.61628328\\
    2460110.366 & 3610.00 & 0.5824948188 & -35.8274029\\
    2460111.356 & 2400.00 & 0.6017307135 & -33.83848388\\
    2460111.365 & 3610.00 & 0.6019237084 & -33.95189577\\
    2460111.398 & 3600.00 & 0.6025582666 & -33.68051113\\
    2460111.408 & 1800.00 & 0.6027464991 & -34.29371441\\
    2460112.348 & 6000.00 & 0.6210390984 & -32.77752988\\
    2460112.358 & 2400.00 & 0.6212307569 & -32.93127401\\
\bottomrule
\end{tabular*}
\end{table*}
\end{center}

\begin{center}
\begin{table*}[t]%
\centering
   \caption{RV measurments of $\alpha$ Dra.\label{table:measurments_vr_5}}%
\tabcolsep=0pt%
\begin{tabular*}{500pt}{@{\extracolsep\fill}lccc}
\toprule
    \textbf{Date of obs. (JD)} &\textbf{Time exposure (s)} & \textbf{Phase} & \textbf{RV (km s$^{-1}$)} \\
\midrule
    2460112.368 & 3610.00 & 0.6214187746 & -32.735841\\
    2460112.369 & 2727.00 & 0.6214385823 & -31.97433211\\
    2460119.37 & 3600.00 & 0.7575902432 & -12.46898018\\
    2460119.371 & 6000.00 & 0.7576076899 & -13.23889327\\
    2460119.471 & 3000.00 & 0.7595485972 & -13.01438483\\
    2460120.384 & 3610.00 & 0.7773140248 & -7.738110698\\
    2460120.388 & 4800.00 & 0.7773852255 & -7.664916784\\
    2460120.461 & 2700.00 & 0.7788028425 & -9.082760194\\
    2460121.373 & 3000.00 & 0.7965493607 & -4.614894105\\
    2460121.377 & 5400.00 & 0.7966172704 & -3.578299171\\
    2460123.369 & 4200.00 & 0.8353539258 & 6.49992198\\
    2460123.377 & 600.00 & 0.8355105014 & 6.804137656\\
    2460123.417 & 2700.00 & 0.836297181 & 9.612644255\\
    2460124.612 & 3600.00 & 0.8595284119 & 17.96535214\\
    2460125.376 & 3600.00 & 0.8743884004 & 21.28394207\\
    2460126.678 & 1206.00 & 0.8997137587 & 31.54653144\\
    2460127.672 & 1067.00 & 0.9190382081 & 39.67144548\\
    2460129.359 & 3600.00 & 0.9518475229 & 49.01371619\\
    2460131.371 & 3600.00 & 0.990990334 & 48.9259388\\
    2460131.715 & 1206.00 & 0.9976775277 & 46.83566524\\
    2460132.405 & 3600.00 & 0.01108795939 & 41.70903937\\
    2460134.364 & 3600.00 & 0.04919255991 & 19.4811746\\
    2460134.38 & 6000.00 & 0.04950295517 & 18.47828099\\
    2460136.355 & 3600.00 & 0.08790710731 & -3.172568522\\
    2460136.378 & 3600.00 & 0.08835664233 & -5.521583602\\
    2460136.693 & 1387.00 & 0.09447579285 & -5.535093623\\
    2460138.354 & 3600.00 & 0.1267852754 & -19.40861455\\
    2460138.371 & 4212.00 & 0.127112825 & -18.7555702\\
    2460139.359 & 3600.00 & 0.14632167 & -25.03969834\\
    2460139.363 & 6000.00 & 0.1464110456 & -22.04977776\\
    2460140.37 & 6000.00 & 0.1659894147 & -29.51747143\\
    2460140.399 & 3600.00 & 0.1665549387 & -29.89678963\\
    2460141.36 & 1202.00 & 0.1852459856 & -32.58759318\\
    2460141.379 & 2400.00 & 0.1856102923 & -34.25017662\\
    2460147.354 & 3600.00 & 0.3018067464 & -43.78766946\\
    2460147.382 & 4500.00 & 0.3023587602 & -43.84135692\\
    2460147.521 & 4200.00 & 0.3050578616 & -43.22653346\\
    2460148.352 & 4200.00 & 0.3212249723 & -44.5448159\\
    2460155.349 & 3000.00 & 0.4573001105 & -43.04371859\\
    2460168.392 & 1800.00 & 0.710955597 & -21.34656808\\
\bottomrule
\end{tabular*}
\end{table*}
\end{center}

\end{document}